\documentclass[5p]{elsarticle}
\usepackage{mathptmx}
\usepackage{flushend}
\usepackage[T1]{fontenc}
\usepackage[utf8]{inputenc}
\usepackage[english]{babel}

\usepackage[colorlinks=true, allcolors=blue, bookmarksopen=true]{hyperref}
\usepackage{amsmath}
\usepackage{textcomp}
\usepackage{amsmath, amsfonts}
\usepackage{subfig}
\usepackage[skip=2pt]{caption} 
\usepackage{xcolor}
\usepackage{setspace}
\biboptions{sort&compress}

\graphicspath{{./Figuras/}}

\begin{document}

\begin{frontmatter}

\journal{Materialia}

\title{Elastic anisotropy and thermal properties of extended linear chain compounds MV$_2$Ga$_4$ (M = Sc, Zr, Hf) from ab-initio calculations}

\author{P. P. Ferreira}
\author{T. T. Dorini}
\author{F. B. Santos}
\author{A. J. S. Machado}
\author{L. T. F. Eleno\corref{cor}}
\ead{luizeleno@usp.br}
\cortext[cor]{Corresponding author: Tel. +55 12 3159 9810}
\address{Lorena School of Engineering, University of S{\~a}o Paulo (EEL--USP), Materials Engineering Department (Demar), Lorena--SP, Brazil}

\begin{abstract}

MV$_2$Ga$_4$ (M = Sc, Zr, Hf) compounds belong to an emerging class of materials showing a unique combination of unusual superconducting behavior with extended linear chains in the crystal structure. In order to gain insights {into} its mechanical and thermal properties, we have performed first-principles electronic-structure calculations in the framework of the Density Functional Theory (DFT). From the calculated second-order elastic constants, we have systematically shown that the extended linear vanadium chain substructures indeed give rise to an anisotropic regime in the elastic and mechanical moduli. The high density of valence and conduction electrons along the linear vanadium chains leads to a directional dependence of the reciprocal linear compressibility, Young's modulus and shear modulus. Poisson's ratio for several elongation directions is also drastically affected by the presence of extended V chains. If the elongation is along the V chains, all compounds exhibit {practically} the same Poisson ratio in directions perpendicular to it, further highlighting the importance of the V chains to the mechanical properties. Moreover, based on our results, we have discussed the possible consequences of the elastic anisotropy on the superconducting properties of the compounds. Finally, using the Debye-Gr\"uneisen approximation, our calculations of thermal properties show {a good agreement with the available experimental low temperature heat capacity data above the superconducting critical temperature.} 

\end{abstract}


\end{frontmatter}

\section{Introduction}

The discovery of new materials drives the technology to an entirely new plethora of applications and possibilities. To harness their potential, it is essential to develop a fundamental understanding of the basic physics that governs their behavior. One of the most interesting class of materials, with crucial applicabilities, are the superconductors. Since their discovery in 1911 by Kamerlingh Onnes, and the first succesful microscopic theory proposed in the seminal work by Bardeen, Cooper and Schrifer (BCS) in 1957 \cite{Bardeen1957}, superconducting materials have been used in medical, energy and transport applications, electronic devices, quantum interference sensors, and as high-field electromagnets \cite{tomita2003, noe2007, tomsic2007, malozemoff2008, putti2010, werfel2011, schwartz2013}. For a long time, novel types of superconducting materials have been found in contrast to what became known as conventional superconductors, which can be explained by BCS-type signatures. In particular, a new type-II electron-phonon superconductor at 4.1\,K was recently discovered with composition HfV$_2$Ga$_4$ \cite{Santos2018}. This compound is interesting and atypical since it exhibits a possible multiband superconducting effect, which fundamentally consists of the opening of two or more superconducting gaps at the Fermi surface below the critical superconducting temperature ($T_c$) \cite{Zehetmayer2013}. In fact, HfV$_2$Ga$_4$ is one of the first examples of a non-heavy fermion material with double-jumps in the specific heat, and the two-gap superconductivity is supported by first-principle calculations, that show electrons arising from distinct bands in disconnected sheets of the Fermi surface with very different electronic characters \cite{Santos2018, ferreira2018}. In addition, it was theoretically predicted that the ScV$_2$Ga$_4$ compound is presumably another example of a two-band electron-phonon superconductor with an even higher $T_c$ than HfV$_2$Ga$_4$ \cite{ferreira2018}.

Attached to the unusual superconducting behavior that deviates from the more conventional BCS-theory signatures, the arrangement of atoms in the HfV$_2$Ga$_4$ compound is such that V atoms form directional bonds with one another so that the structure, as a whole, achieves extended linear 1D highly-populated (by electrons) V-$d$ chains that command the density of states at the Fermi level ($E_F$). Extended (or ``infinite'', as some authors prefer) linear chain compounds, despite known for quite some time, are a relatively recent topic of interest, having been systematically investigated only in the last few years in the literature \cite{miller2012}. Those compounds generally exhibit useful electrical, magnetic, superconducting, thermal and optical anisotropic properties, as seen, for instance, in Hg$_{3-\delta}$AsF$_6$ and Hg$_{3-\delta}$SbF$_6$ compounds \cite{cutforth1977, chiang1977, koteles1976, peebles1977}. Their anisotropy occurs due to the high density of conduction electrons on the Hg chains, which act as one-dimensional metals \cite{brown1983}. Transition metal chain substructures were also reported in several compounds within the YbMo$_2$Al$_4$-prototype (the same prototype of MV$_2$Ga$_4$ materials, with M = Sc, Zr, Hf), but most efforts were focused on chemical bonding and electronic properties \cite{muts2011, matar2013, gerke2013, tappe2013}. There is, therefore, a lack of information on the effect of the extended linear chains on the mechanical properties of such compounds. Moreover, ab-initio calculations of mechanical properties are a challenging but promissing simulation technique that can be used with advantage to probe the potencial of new materials \cite{Pokluda2015}.

In this work we have shown, using Density Functional Theory (DFT)-based methods, that the extended linear vanadium chains in the MV$_2$Ga$_4$ (M = Sc, Zr, Hf) compounds indeed give rise to an anisotropic regime in the elastic and mechanical moduli. From the calculated second-order elastic constants and the Debye-Gr\"uneisen quasi-harmonic approximation, we have evaluated the mechanical and thermal properties, as well as the directional dependence of linear compressibility, shear and Young's modulus, Poisson's ratio, and sound velocities in the crystal structure. 
%Our results show an excellent agreement with the scanty experimental data available. 
Similar methodologies have been extensively employed in the literature to obtain mechanical and thermal properties for widely different classes of materials \cite{xiao2011, feng2011, feng2012, feng2013, sun2013, gao2014, yang2016}, which point to the importance of the topic and its high potential to promote and motivate future studies. On the same footing, it is possible that other extended linear chain compounds present similar elastic anisotropy trends as the compounds studied here, pointing to a general behavior of this class of materials.

\section{Computational Methods}

\label{sec:methods}

Ab-initio calculations were performed using the Quantum \textsc{Espresso} computational package \cite{Giannozzi2009} within the framework of Density Functional Theory (DFT) in the Kohn-Sham scheme \cite{kohn1965}, employing scalar-relativistic optimized norm-conserving Vanderbilt pseudopotentials \cite{Schlipf2015}. The Exchange and Correlation (XC) functional employed was the Generalized Gradient Approximation (GGA) \cite{gross2013} in the parametrization due to Perdew-Burke-Ernzerhof (PBE) \cite{perdew1996}. We have used a wavefunction energy cut-off of 220\,Ry (1\,Ry $\approx$ 13.6\,eV), and four times that value for the charge density energy cut-off. The Monkhorst-Pack scheme \cite{monkhorst1976} was used for a 1728 $k$-point sampling in the first Brillouin zone. Self-consistent-field (SCF) calculations were carried out using Marzari-Vanderbilt cold smearing \cite{marzari1999} of 0.002\,Ry, and the damped quick-min Verlet ion dynamics algorithm was applied for structural relaxation \cite{arias1992}. All latice parameters and internal degrees of freedom were relaxed in order to guarantee a ground-state convergence of 10$^{-5}$\,Ry in total energy and 0.5\,mRy/$a_0$ ($a_0\approx0.529\,$\AA) for forces acting on the nuclei.
{In a previous work \cite{ferreira2018} we used an all-electron approach, instead of the pseudopotential methodology applied here, in order to obtain the ground-state electronic properties for the HfV$_2$Ga$_4$ and ScV$_2$Ga$_4$ compounds. We found no appreciable differences either in the density of states or in the band structure plots when we compared the two techniques, which we take as an indication of the validity of the present methodology and of the choice of pseudopotentials.}

The full second-order elastic stiffness tensor was obtained using the ElaStic code \cite{elastic}. In the energy approach, the elastic constants result from a set of deformations imposed on the reference ground-state structure, in the framework of the Lagrangian Theory of Elasticity \cite{kantorovich2004}:
\begin{equation}
  c_{\alpha\beta} = \frac{1}{V_0} \left. \frac{\partial^2 U}{\partial \eta_{\alpha} \partial \eta_{\beta}} \right|_{\eta=0},
\end{equation}
where $U$ is the total energy due to the deformation, $V_0$ is the volume of the undeformed ground state structure and $\eta_\alpha$, $\eta_\beta$ are Lagrangian strains, expressed in Voigt notation. The MV$_2$Ga$_4$ compounds crystallize in a body-centered tetragonal structure, having therefore six independent second-order elastic constants, as shown below in matrix form:
\begin{equation}
 \mathbb{C} =
 \begin{pmatrix}
c_{11} & c_{12} & c_{13} & 0 & 0 & 0 \\
c_{12} & c_{11} & c_{13} & 0 & 0 & 0 \\
c_{13} & c_{13} & c_{33} & 0 & 0 & 0 \\
0 & 0 & 0 & c_{44} & 0 & 0\\
0 & 0 & 0 & 0 & c_{44} & 0\\
0 & 0 & 0 & 0 & 0 & c_{66} \\
\end{pmatrix}
\end{equation}
Therefore, in the ElaStic code we have used six different deformation types for each ground-state compound, with a maximum absolute intensity of $\eta_\text{max} = 0.05$ for the Lagrangian strain, and 15 distorted structures with strain intensities between $-\eta_\text{max}$ and $+\eta_\text{max}$ for each deformation type, a total of 90 deformations per compound.

\section{Results and Discussion}

\subsection{Elastic stiffness and compliance tensors}
\label{sec:stiff}

\begin{figure}[t]
\centering
\includegraphics[width=.7\columnwidth]{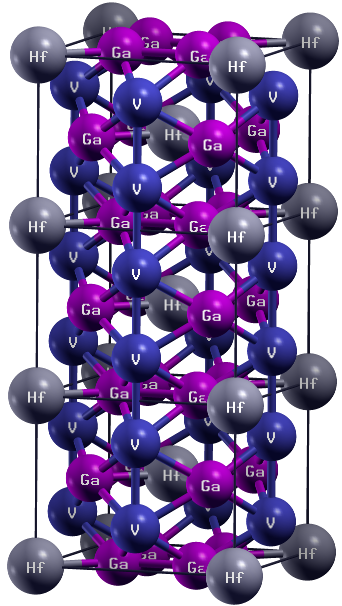}
\caption{Three unit cells of the MV$_2$Ga$_4$ crystal structure stacked along the $c$-direction, exemplified for M = Hf. The V atoms form extendend linear chains in the crystal, oriented paralell to the  $c$-axis.}
\label{fig:structure}
\end{figure}

MV$_2$Ga$_4$ (M = Sc, Zr, Hf) compounds crystallize in the same body-centered tetragonal structure as the YbMo$_2$Al$_4$ prototype (space group $I4/mmm$, \#139, Pearson symbol $tI14$) \cite{fornasini1976}. The crystal structure is such that the M sites, at the $2a$ (0, 0, 0) Wyckoff positions, are surrounded by V and Ga sites at $4d$ (0, 1/2, 1/4) and $8h$ (0.303, 0.303., 0), respectively, in a cage-like structure, in such a way that the vanadium sites form extended linear chains in the $c$-direction. A stacking of three unit cells along the $c$ axis is schematically ilustrated in Figure \ref{fig:structure}. In the present work, the initial crystal structure was built based on experimental crystallographic data \cite{fornasini1976, Santos2018}, and then optimized as described in Sec. \ref{sec:methods}. The calculated lattice parameters and the $8h$ degree of freedom ($x_\text{Ga}$) are shown in Table \ref{tab:lat-par}. The slight dissimilarity of at most 1\% in the optimized lattice parameters and atomic positions in comparison with experimental data reinforce the validity of the computational methods employed.

\begin{table}[b]
	\caption{Calculated lattice parameters and optimized $8h$ ($x_\text{Ga}$) atomic position for the MV$_{2}$Ga$_{4}$ (M = Sc, Zr, Hf) body-centered tetragonal compounds, compared to experimental values \cite{fornasini1976, Santos2018}.}
	\label{tab:lat-par}
	\centering
	\begin{tabular}{ccccc}
		\hline
		& & $a$ (\AA) & $c$ (\AA) & $x_\text{Ga}$\\
		\hline
		ScV$_2$Ga$_4$ & calc. & 6.421 & 5.155 & 0.3003 \\[-1mm]
		& exp. & 6.432 & 5.216 & 0.3030 \\[2mm]
		ZrV$_2$Ga$_4$ & calc. & 6.379 & 5.140 & 0.3026 \\[-1mm]
		& exp. & 6.462 & 5.207 & 0.3030 \\[2mm]
		HfV$_2$Ga$_4$ & calc. & 6.458 & 5.206 & 0.3030 \\[-1mm]
		& exp. & 6.432 & 5.190 & 0.3030 \\				
		\hline
	\end{tabular}
\end{table}

%The elastic stiffness constant $c_{\alpha\beta}$ measures the tendency of a solid to deform when a external stress is applied to it, and provide a set of valuable informations about bonding, mechanical and thermal properties. 
%The number of independent parameters required to fully describe $c_{\alpha\beta}$ depends on the crystallographic symmetry of the system. 

The six calculated second-order elastic constants necessary to specify the elastic stiffness tensor for a body-centered tetragonal crystal are listed in Table \ref{tab:cmatrix}. The related elastic compliance constants $s_{\alpha\beta}$ are presented in Table \ref{tab:smatrix}, obtained by matrix inversion of the stiffness tensor:
\begin{equation}
 \sigma_\alpha = c_{\alpha\beta} \varepsilon_\beta \qquad \Leftrightarrow \qquad \varepsilon_\alpha = s_{\alpha\beta} \sigma_\beta
\end{equation}
in which $1 \le \alpha,\,\beta \le 6$ and, as in the rest of the work, Einstein's tensorial implicit, equal indices sum notation was employed.
It is seen that the elastic constants in Table \ref{tab:cmatrix} do not obey the Cauchy conditions $c_{12}=c_{66}$ and $c_{13}=c_{44}$, a indication that the strenght does not come from central forces. This, naturally, is a consequence of the symmetry of the structure, since, with the exception of the the M sites (occupied by Sc, Zr or Hf), the atomic positions are not centers  of inversion \cite{born}.

For tetragonal crystals, the elastic stability can be verified using the four necessary and sufficient conditions established by \citet{mouhat2014}:
\begin{align}
	c_{11} > \left|c_{12}\right|\,, \nonumber \\
	2c_{13}^2 < c_{33}(c_{11} + c_{12})\,, \\
	c_{44} > 0\,, \qquad
	c_{66} > 0\,. \nonumber
\end{align}
All of the conditions listed above are fulfilled by the elastic constants in Table \ref{tab:cmatrix}, a result that point to the mechanical stability of the MV$_2$Ga$_4$ (M = Sc, Zr, Hf) compounds. 

The elastic constants $c_{\alpha\beta}$ also provide straightforward information about the elastic response to different applied stress conditions in single crystals. For instance, $c_{11}$ and $c_{33}$ represent the resistance to unixial deformation along the [100] and [001] directions, respectively, while $c_{44}$ is related to the resistence to a pure shear deformation on $(hk0)$ planes. One can verify in Table \ref{tab:cmatrix} that ScV$_2$Ga$_4$ and HfV$_2$Ga$_4$ are slightly more resistent under uniaxial stress along the $z$ direction than along the $x$ or $y$ axis, while ZrV$_2$Ga$_4$ has practically the same resistance in both these orientations. ZrV$_2$Ga$_4$ also has the largest $c_{44}$ elastic constant, which indicates an appreciable shear modulus in the plane normal to the extended vanadium chains, as will be seen in more detail in section \ref{sec:anis}.

\begin{table}
	\caption{Calculated second-order elastic stiffness constants (in GPa) of the body-centered tetragonal MV$_2$Ga$_4$ (M = Sc, Zr, Hf) compounds.}
	\label{tab:cmatrix}
% 	\small
	\centering
	\begin{tabular}{ccccccc}
		\hline
		& $c_{11}$ & $c_{12}$ & $c_{13}$ & $c_{33}$ & $c_{44}$ & $c_{66}$ \\
		\hline
		ScV$_2$Ga$_4$ & 235.6 & 121.3 & 65.0 & 261.4 & 89.1 & 137.5 \\[2mm]
		ZrV$_2$Ga$_4$ & 288.7 & 133.6 & 76.7 & 283.0 & 102.5 & 157.9 \\[2mm]
		HfV$_2$Ga$_4$ & 236.9 & 120.7 & 64.9 & 242.2 & 80.2 & 131.6 \\				
		\hline
	\end{tabular}
\end{table}

\begin{table}
	\caption{Elastic compliance constants (in TPa$^{-1}$) of the MV$_2$Ga$_4$ (M = Sc, Zr, Hf) compounds, calculated from the full second-order elastic stiffness tensor.}
	\label{tab:smatrix}
	\centering
	\begin{tabular}{ccccccc}
		\hline
		& $s_{11}$ & $s_{12}$ & $s_{13}$ & $s_{33}$ & $s_{44}$ & $s_{66}$ \\
		\hline
		ScV$_2$Ga$_4$ & 5.91 & $-$2.83 & $-$0.77 & 4.21 & 11.23 & 7.27 \\[2mm]
		ZrV$_2$Ga$_4$ & 4.54 & $-$1.91 & $-$0.71 & 3.92 & 9.76 & 6.33 \\[2mm]
		HfV$_2$Ga$_4$ & 5.85 & $-$2.75 & $-$0.83 & 4.57 & 12.47 & 7.60 \\		
		\hline
	\end{tabular}
\end{table}

\subsection{Polycrystal mechanical properties}

\begin{table*}[t]
	\centering
	\caption{Calculated Bulk modulus ($B$), shear modulus ($G$), Young's modulus ($E$) and Poisson's ratio ($\nu$) for the MV$_2$Ga$_4$ (M = Sc, Zr, Hf) compounds according to the Voigt-Reuss-Hill approximation. All values are in GPa (except the adimentional quantities).}
	\label{tab:mechanical}
	\footnotesize
	\begin{tabular}{cccccccccccccc}
		\hline
		& $B_V$ & $B_R$ & $B_H$ & $G_V$ & $G_R$ & $G_H$ & $B/G$ & $E_V$ & $E_R$ & $E_H$ & $\nu_V$ & $\nu_R$ & $\nu_H$ \\
		\hline		
		ScV$_2$Ga$_4$ & 137.24 & 136.91 & 137.08 & 95.21 & 87.83 & 91.52 & 1.498 & 231.99 & 217.07 & 224.58 & 0.22 & 0.24 & 0.23\\[2mm]
		ZrV$_2$Ga$_4$ & 159.37 & 158.08 & 158.73 & 110.80 & 105.01 & 107.91 & 1.471 & 269.87 & 257.93 & 263.92 & 0.22 & 0.23 & 0.22\\[2mm]
		HfV$_2$Ga$_4$ & 135.23 & 134.26 & 134.75 & 89.43 & 83.15 & 86.29 & 1.562 & 219.83 & 206.77 & 213.33 & 0.23 & 0.24 & 0.24\\
		\hline
	\end{tabular}
\end{table*}

Once the set of independent elastic stiffness and compliance constants for single crystals are known, we can calculate the mechanical properties of a polycristalline aggregate using the Voigt-Reuss-Hill (VRH) approximation. In the Voigt approach an uniform strain is assumed, while Reuss considers an uniform applied stress. \citet{hill1952} has shown that the measured moduli for the agregate always lie in between the Voigt and Reuss procedures, therefore proposing the bulk ($B$) and shear ($G$) moduli of the polycrystalline material as the arithmetic average of Voigt (upper) and Reuss (lower) bounds:
\begin{align}
	B_H &= \dfrac{1}{2}(B_V + B_R), \\
	G_H &= \dfrac{1}{2}(G_V + G_R).
\end{align}
with $B_V$, $G_V$, $B_R$, and $G_R$, for body-centered tetragonal structures, given respectively by
\begin{equation}
 B_V = \frac{2c_{11}+c_{33}+2(c_{12}+2c_{13})}{9} 
 \end{equation}
 \begin{equation}
 G_V = \frac{2c_{11}+c_{33}-(c_{12}+2c_{13})+3(2c_{44}+c_{66})}{15} 
 \end{equation}
 \begin{equation}
 B_R^{-1} = 2s_{11}+s_{22}+2(s_{12}+2s_{13}) 
 \end{equation}
 \begin{equation}
 G_R^{-1} = \frac{4(2 s_{11} + s_{33})-(s_{12} + 2 s_{13})+3(2 s_{44} + s_{66})}{15}
\end{equation}

Furthermore, Young's modulus ($E$) and Poisson's ratio ($\nu$) are obtained in terms of the bulk and shear modulus using the isotropic theory of elasticity \cite{timoshenko2008}:
\begin{align}
	E &= \dfrac{9BG}{3B + G}, \\
	\nu &= \dfrac{3B - 2G}{2(3B+G)}.
\end{align} 

Following the VRH approximation, the mechanical properties according to Voigt, Reuss and Hill approaches were calculated and are listed in Table \ref{tab:mechanical}. We verify that all MV$_2$Ga$_4$ compounds have the same order of magnitude regarding mechanical moduli, indicating that the element on the 2a site in the structure has only a weak influence on the mechanical properties when we take the overall scenario into account. The larger difference between them is observed in ZrV$_2$Ga$_4$, which has the largest values for bulk, shear and Young's moduli. The obtained bulk modulus, which indicates the resistence of a solid to a hydrostatic external pressure, reveals that ZrV$_2$Ga$_4$ has the most significant resistence to volume change under pressure.

The $B/G$ ratio can be used to predict the ductility of a material \cite{pugh1954}, and has been extensivily used for this purpose in the recent literature \cite{xiao2010, feng2013, duan2014, gao2014}. A $B/G$ ratio higher than the critical value 1.75 indicates a significant ductile material, while $B/G<1.75$ usually means that the material is brittle. As shown in Table \ref{tab:mechanical}, all the compounds are essentially brittle, consistent with Poisson's ratio values smaller than 0.26. It is known that the Poisson's ratio is a good indicative of the degree of covalent bonds, telling also in advance about the ductility and brittleness of a polycrystalline material \cite{chen2011}.

\subsection{Anisotropy of elastic properties}
\label{sec:anis}

In several materials, the degree of anisotropy in the single crystal elastic properties is essential to unravel the macroscopic behavior of a solid, even in the polycristalline state \cite{cazzani2005}.  As previously discussed, extended linear chain compounds usually present strong anisotropy. Therefore, it is reasonable that mechanical moduli also comes up with an outward directional dependence directly related to the elements of the stiffness tensor. Given its importance to applications, considerable efforts have been made to develop parameters able to quantify the extent of anisotropy in elasticity \cite{zener1948, chung1967, ledbetter2006}. Most recently, Ranganathan et. al \cite{ranganathan2008} proposed a new universal index to quantify a single crystal elastic anisotropy:
\begin{align}
	A^{U} = 5\frac{G_V}{G_R} + \frac{B_V}{B_R} - 6.
\end{align}
$A^{U}$ is identically zero for locally isotropic single crystals, and deviations from zero point out the degree of elastic anisotropy of the material. Using the values in Table \ref{tab:mechanical} we see that ZrV$_2$Ga$_4$, despite possessing the highest mechanical moduli values, has the smallest universal anisotropy index, around 0.28. HfV$_2$Ga$_4$ presents an index equal to 0.38, revealing a considerable directional dependence. The strongest elastic anisotropy, thought, is reached by ScV$_2$Ga$_4$, which has a value of 0.42 for $A^U$. The main contribution for  $A^U$ in the MV$_2$Ga$_4$ compounds comes from the shear moduli, taking into account the large difference between the $G_V$ and $G_R$ values according to Voigt and Reuss approximations, as seen in Table \ref{tab:mechanical}.

\begin{figure}
  \centering
  \includegraphics[width=.8\columnwidth]{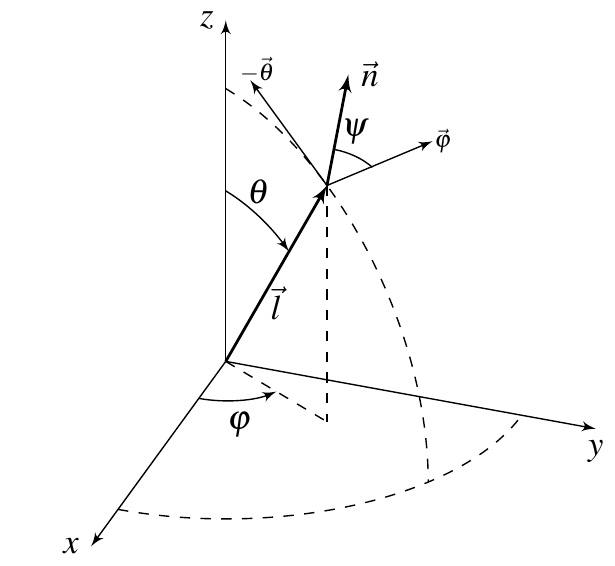}
 \caption{Definition of angles and directions used in the present work. The angles $\varphi$ and $\theta$ are, respectively, the azimuthal and polar angles in spherical coordinates that define the tension (or elongation) direction $\vec l$. Direction $\vec n$, normal to $\vec l$, forms an angle $\psi$ with the unit vector $\vec \varphi$.}
 \label{fig:angles}
\end{figure}

At the same time, the directional dependence of the reciprocal linear compressibility ($B_c$) and Young's modulus ($E$) can be most easily visualized in the form of 3D spherical plots. In the full fourth-rank tensorial notation, these are given, respectively, by \cite{Hayes1998, nye1985}
\begin{align}
 B_c^{-1}(\vec l) = s_{ijkk} l_{i} l_{j} \label{eq:B--4th}\\
 E^{-1}(\vec l) = s_{ijkm} l_{i} l_{j} l_{k} l_{m} \label{eq:E--4th}
\end{align}
in which $1 \le i,\, j,\, k,\, m \le 3$ and $\vec l = (l_1,\, l_2,\,l_3)$ is a unit vector defining the tension direction. For instance, we can adopt 
\begin{equation}
\vec l = 
 \begin{pmatrix}
  l_1 \\ l_2 \\ l_3
 \end{pmatrix}
=
\begin{pmatrix}
  \cos \varphi \sin\theta \\
  \sin \varphi \sin \theta \\
  \cos \theta
 \end{pmatrix}
\end{equation}
using the usual azimuthal ($\varphi$) and polar ($\theta$) angles in spherical coordinates shown schematically in Figure \ref{fig:angles}.

We take the opportunity to make a remark about the reciprocal linear compressibility coefficient $B_c$, since there seems to be a certain degree of confusion in the literature regarding this material property. Some authors equate it to the bulk modulus of the single crystal along a given direction. This interpretation lacks physical meaning, since the bulk modulus is essentially a non-directional quantity, different from the linear compressibility ($B_c^{-1}$), which is indeed a directional property that tells the linear contraction along a given direction resulting from a hydrostatic external pressure \cite{nye1985}. That explains the fact that the ``bulk modulus'' spherical plots of some authors show values around three times higher than tabulated values found in the same reference (see refs. \citenum{feng2012, sun2013, gao2014}, for instance).

Returning to Eqs.\,(\ref{eq:B--4th}) and (\ref{eq:E--4th}) applied to a body-centered tetragonal system and reverting to Voigt's notation \cite{nye1985}, we can rewrite $B_c(\vec l)$ and $E(\vec l)$ along a given direction $\vec l$ as
\begin{align}
B_c^{-1}(\vec l) & = (s_{11} + s_{12} + s_{13})\left(l_1^2 + l_2^2\right) + (2s_{13} + s_{33})l_3^2 \,, \\
E^{-1}(\vec l) &= s_{11}\left(l_1^4 + l_2^4\right) + s_{33}l_3^4 + \nonumber \\
	&  + (2s_{13} + s_{44})\left(l_1^2 + l_2^2 \right) l_3^2  + (2s_{12} + s_{66})l_1^2 l_2^2\,. 
\end{align}

\begin{figure*}[t]
	\centering
	\subfloat[][ScV$_2$Ga$_4$]{\includegraphics[width=.33\textwidth]{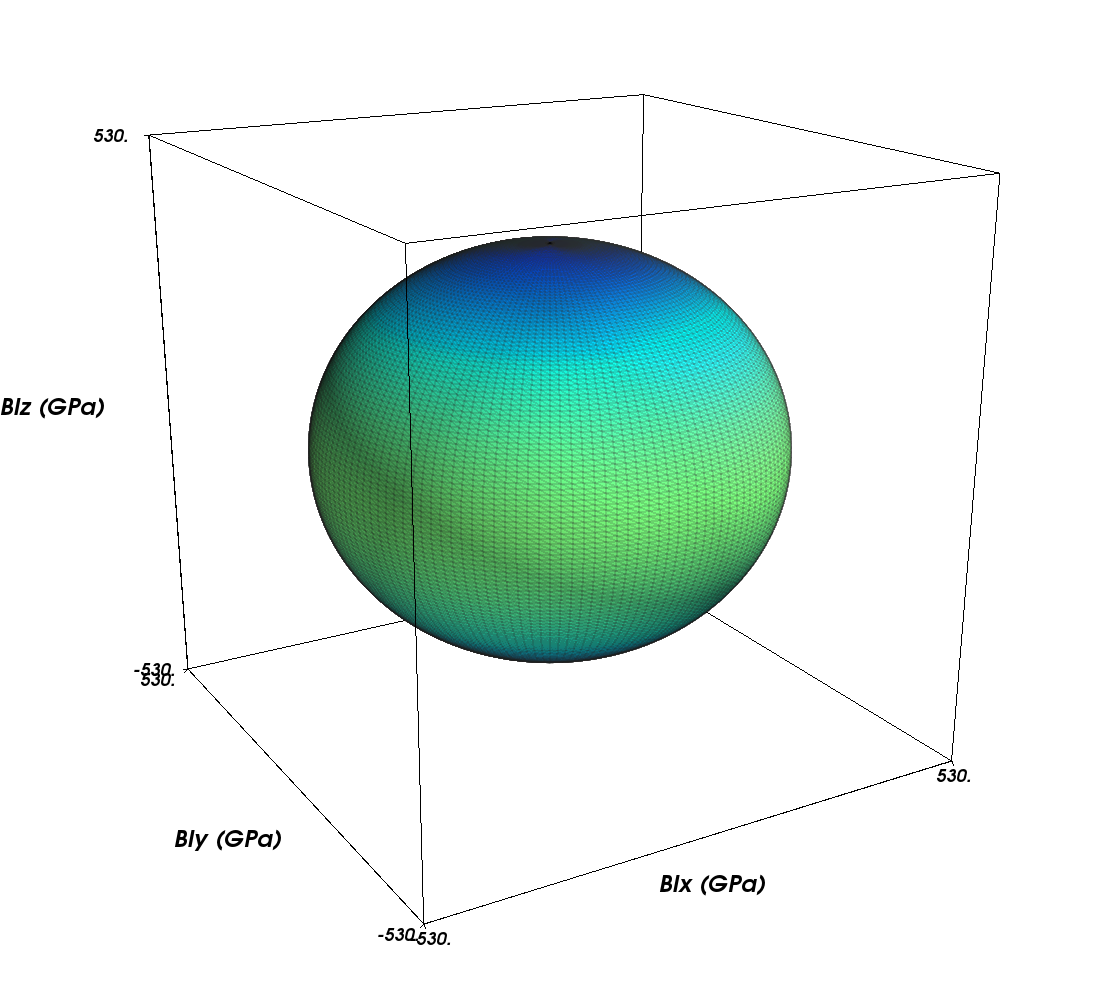}}
	\subfloat[][ZrV$_2$Ga$_4$]{\includegraphics[width=.33\textwidth]{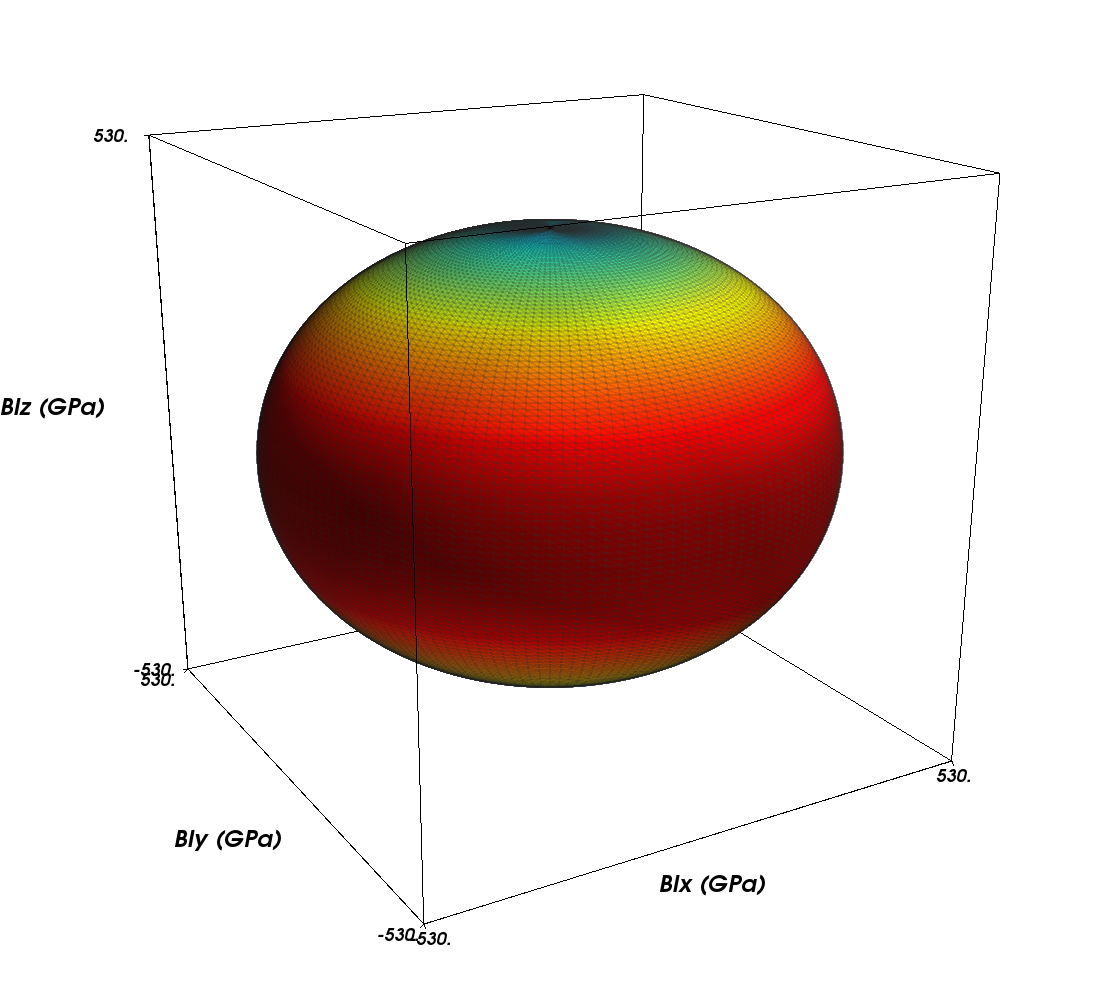}}
	\subfloat[][HfV$_2$Ga$_4$]{\includegraphics[width=.33\textwidth]{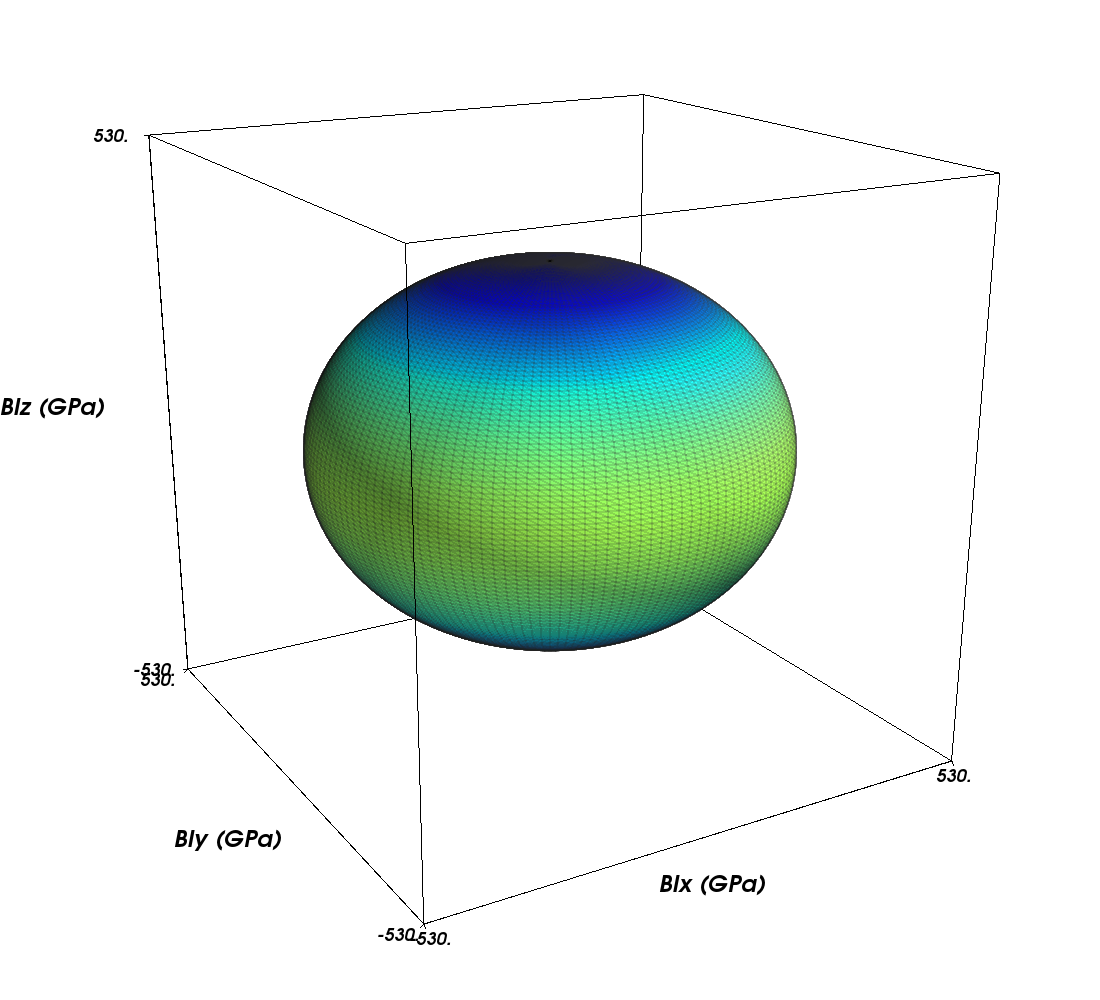}} \\
	\includegraphics[width=.4\textwidth]{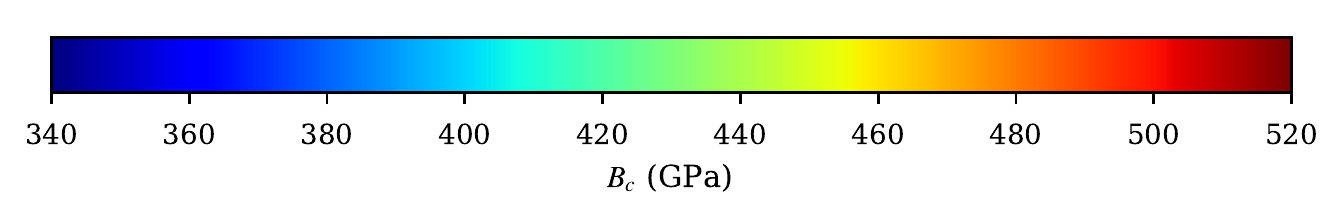}\\
	\subfloat[][ScV$_2$Ga$_4$]{\includegraphics[width=.33\textwidth]{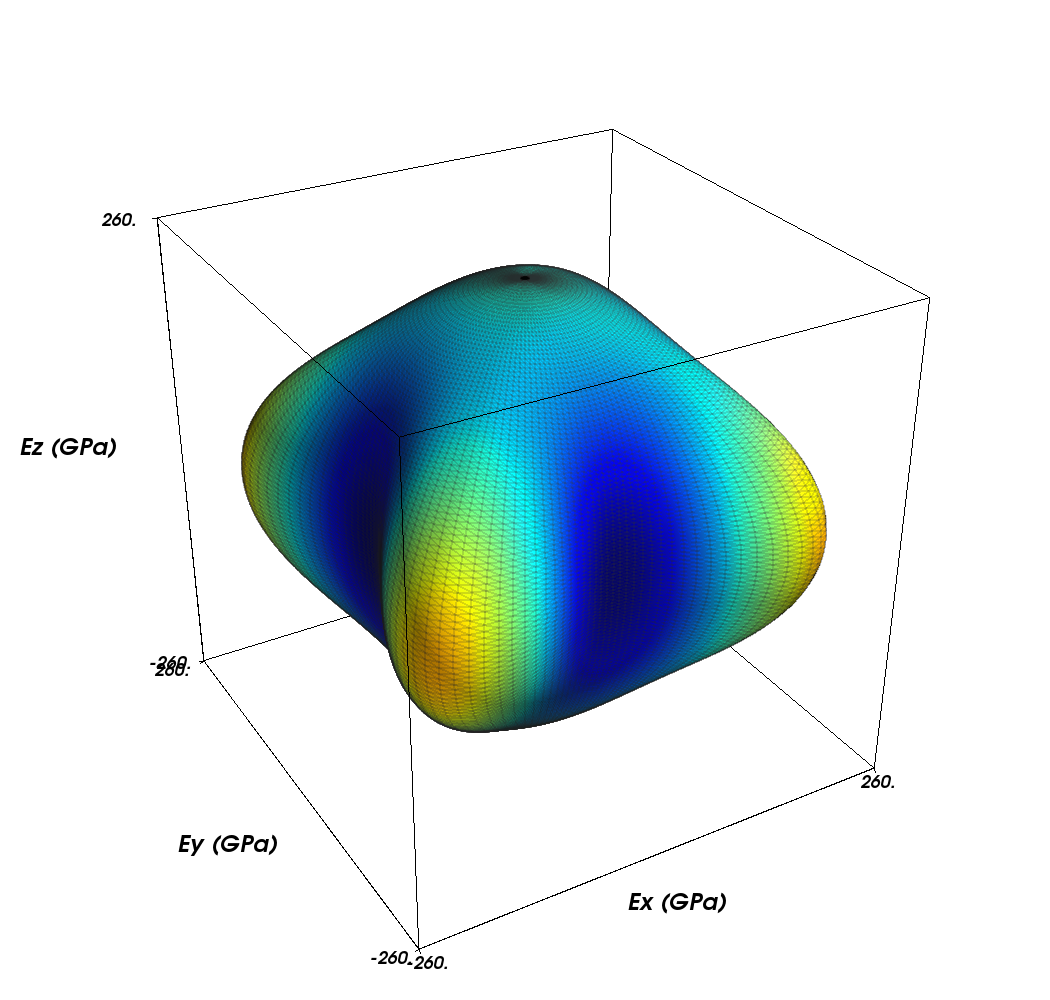}}
	\subfloat[][ZrV$_2$Ga$_4$]{\includegraphics[width=.33\textwidth]{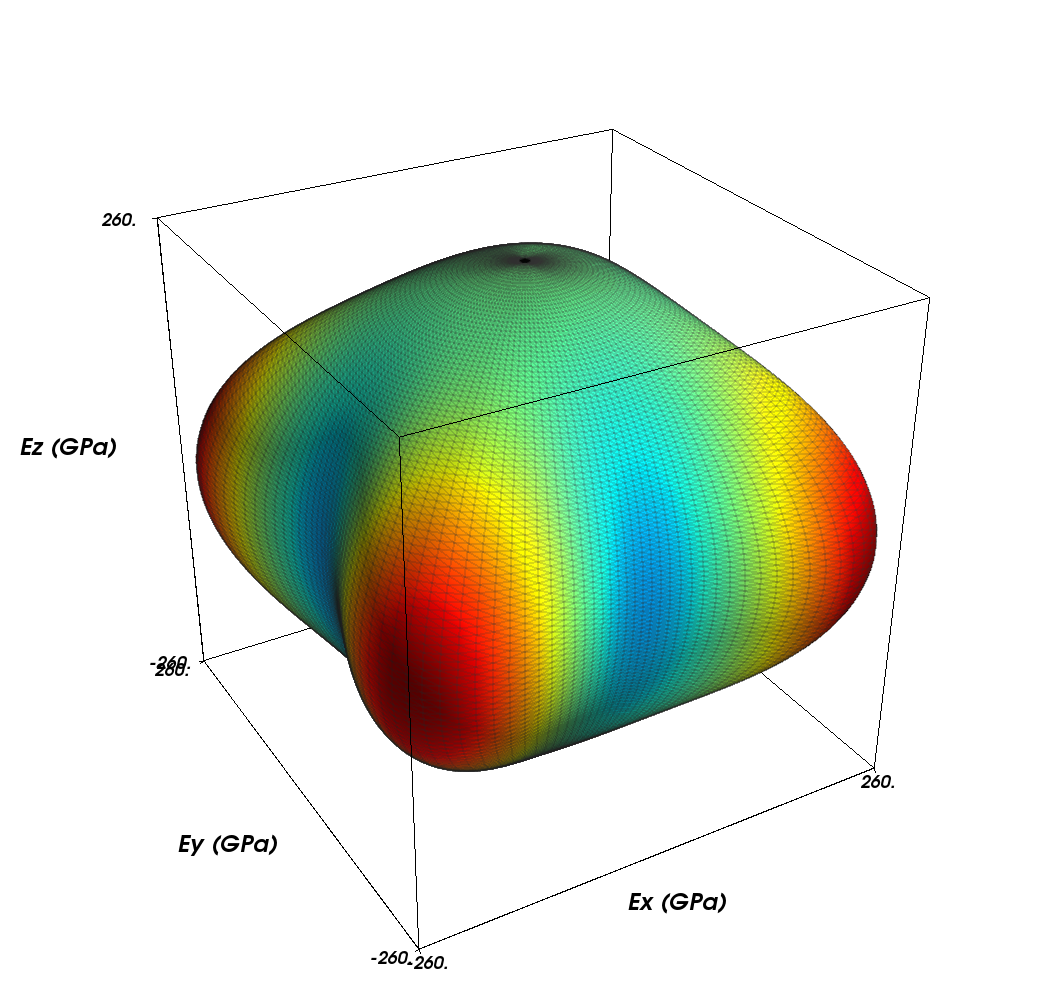}}
	\subfloat[][HfV$_2$Ga$_4$]{\includegraphics[width=.33\textwidth]{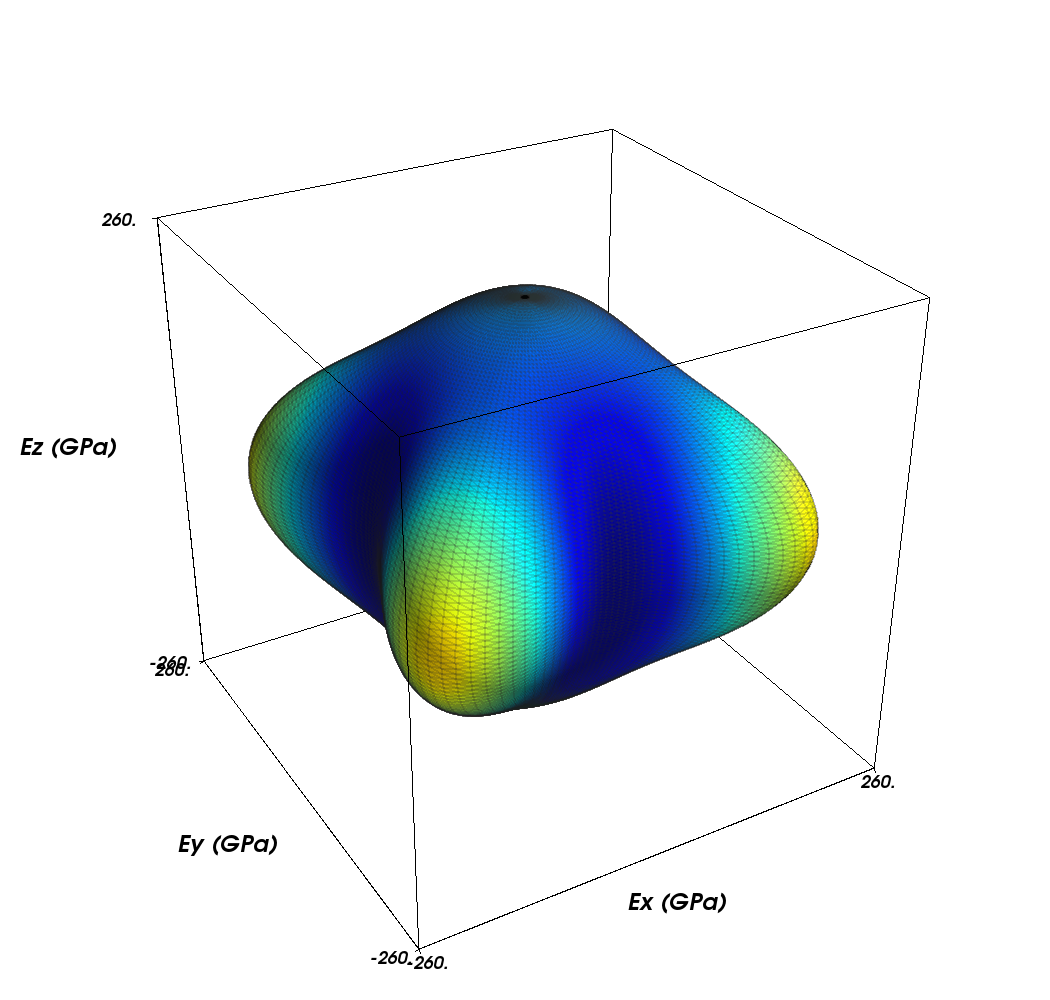}} \\
	\includegraphics[width=.4\textwidth]{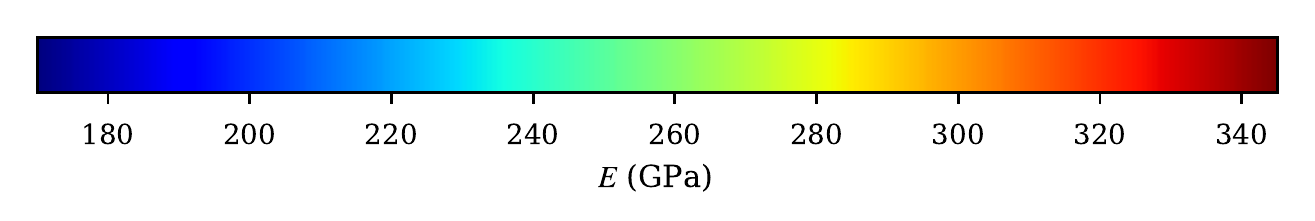}
	\caption{Surface spherical plots showing the directional depence of the reciprocal linear compressibility $B_c$ (a-c) and Young's modulus $E$ (d-f) for the MV$_2$Ga$_4$ (M = Sc, Zr, Hf). Values in GPa.}
	\label{fig:mechanical_anisotropy}
\end{figure*}

The resulting spherical plots for the reciprocal linear compressibility and Young's modulus as a function of the crystallographic orientation are shown in Figure \ref{fig:mechanical_anisotropy}(a-f). In the plots, colours indicate the magnitude of the moduli, going in ascending scale from blue to red. The reciprocal linear compressibility plots show only a weak anisotropy, presenting a quasi-spherical profile. As with any chain, the extended linear V chains are weaker against compression and, therefore, the $B_c$ plots are slighty flattened in the [001] direction, as seen in Figures \ref{fig:mechanical_anisotropy}(a-c).

A strong elastic anisotropy, on the other hand, is observed from the Young's modulus plots. The degree of deviation from the isotropic spherical shape points to the directionality of the material properties. A large resistence to elastic deformation is observed in the [110] direction, consistent with the fact that the (001) plane in the MV$_2$Ga$_4$ compounds has the higher planar density in the crystal, mainly due to the Hf--Ga bonds along [110]. In addition, as demonstrated by Rovati and Cazzani \cite{cazzani2005}, the elastic response of tetragonal solids is intrinsicaly restricted by symmetry considerations and can be classified in 12 different and well-defined classes. As a consequence, it seems to be a fairly common feature in a great number of real single crystals with tetragonal symmetry that the Young's modulus in directions within the (110) plane assumes higher values \cite{cazzani2005}. However, the Young's modulus is also high along the [001] direction for the MV$_2$Ga$_4$ compounds, despite having a lower planar density at the (hk0) planes. The reason for this is directly linked to the presence of extended linear vanadium chains along the $c$-direction. For instance, it is easy to realise from the spherical plots that the anisotropy of Young's modulus along the [001] direction in ZrV$_2$Ga$_4$ is less prominent than that of HfV$_2$Ga$_4$ and ScV$_2$Ga$_4$. This is in agreement with our previous analysis of the  universal elastic anisotropy index, higher for ScV$_2$Ga$_4$ and HfV$_2$Ga$_4$ and lower for ZrV$_2$Ga$_4$. Therefore, we can conclude that the extended linear chains indeed play a fundamental role in the mechanical properties and anisotropy indices.

\begin{figure*}[t]
	\centering
	\subfloat[][]{\includegraphics[width=.33\textwidth]{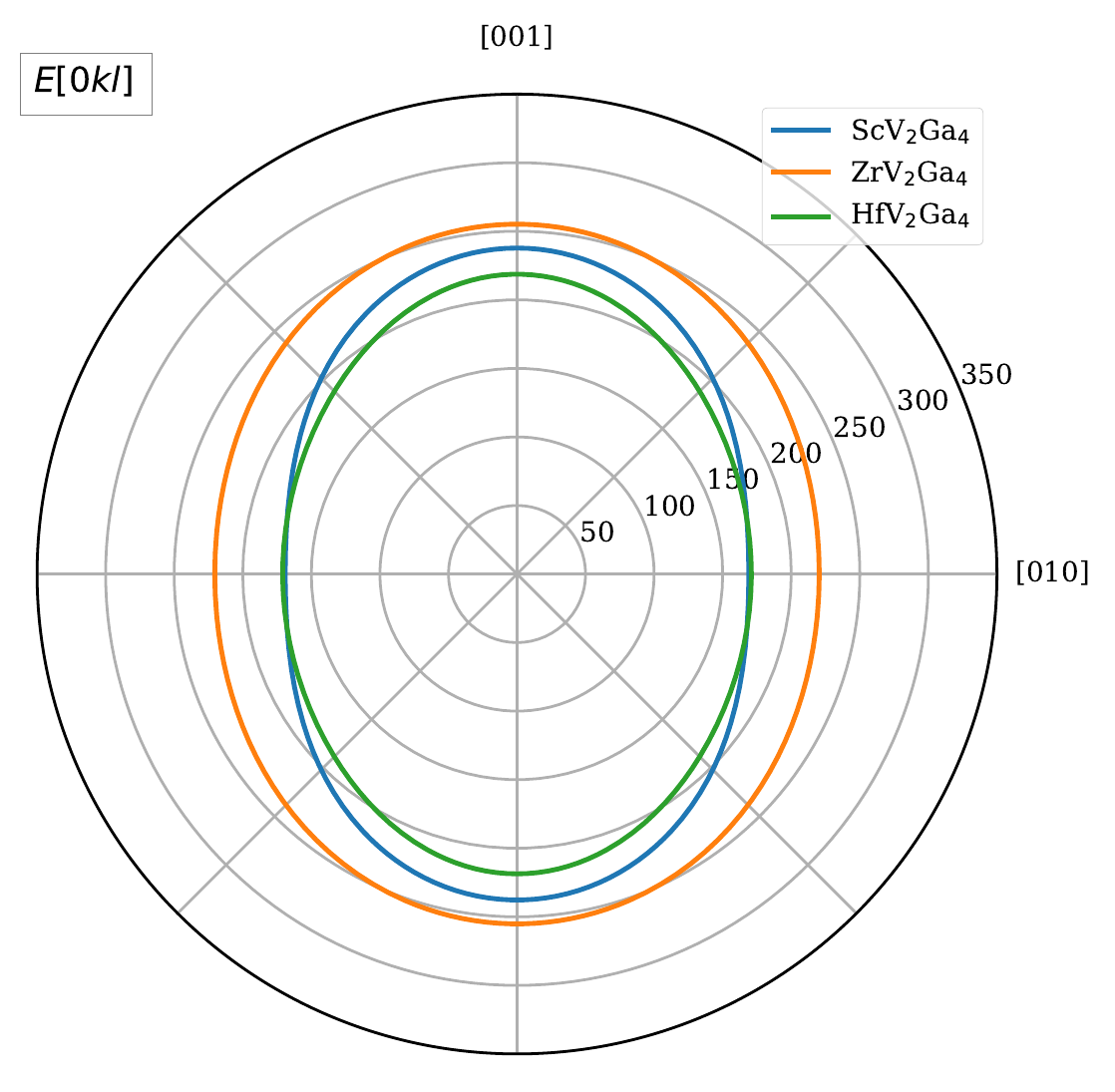}}
	\subfloat[][]{\includegraphics[width=.33\textwidth]{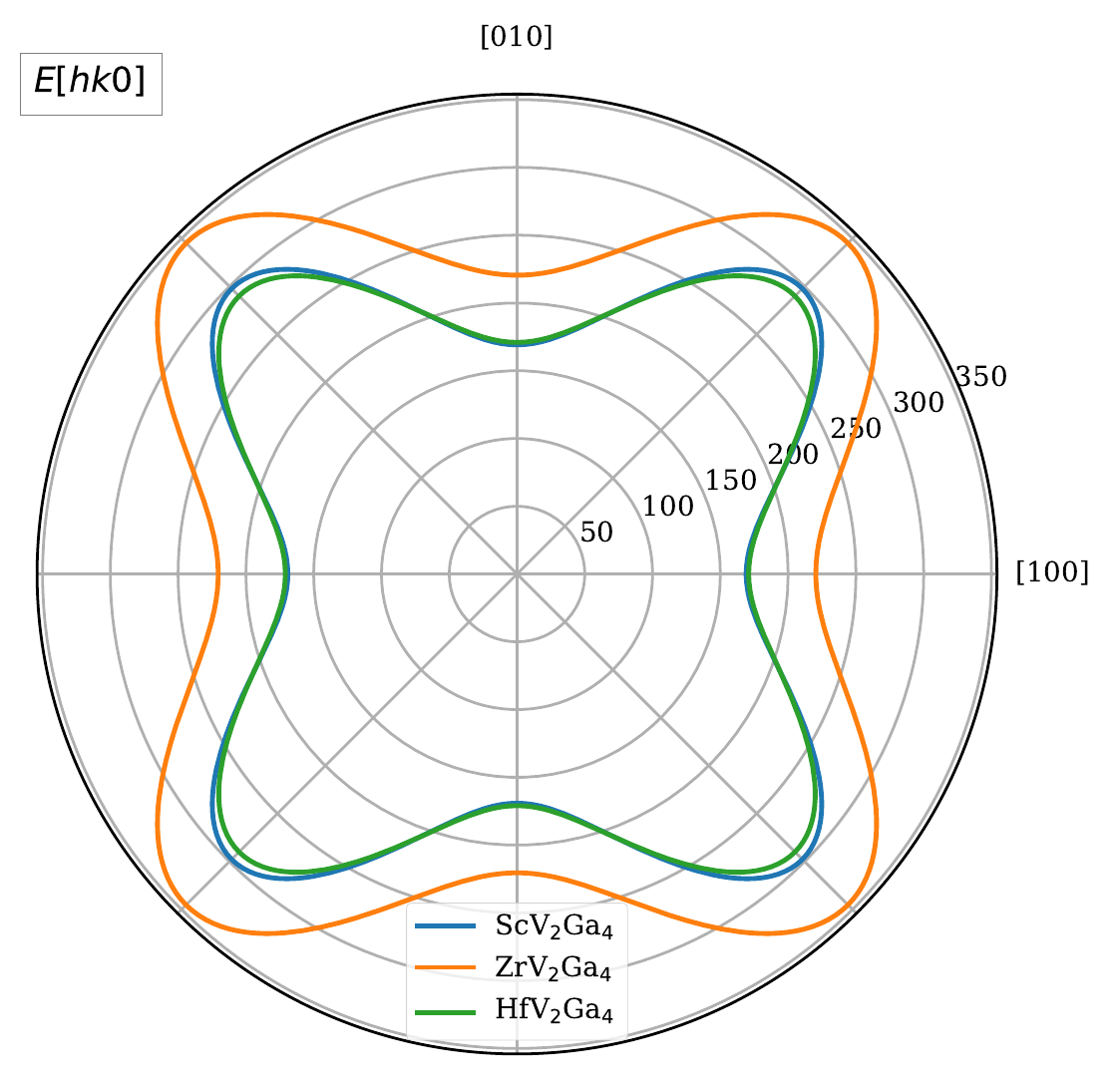}}
	\subfloat[][]{\includegraphics[width=.33\textwidth]{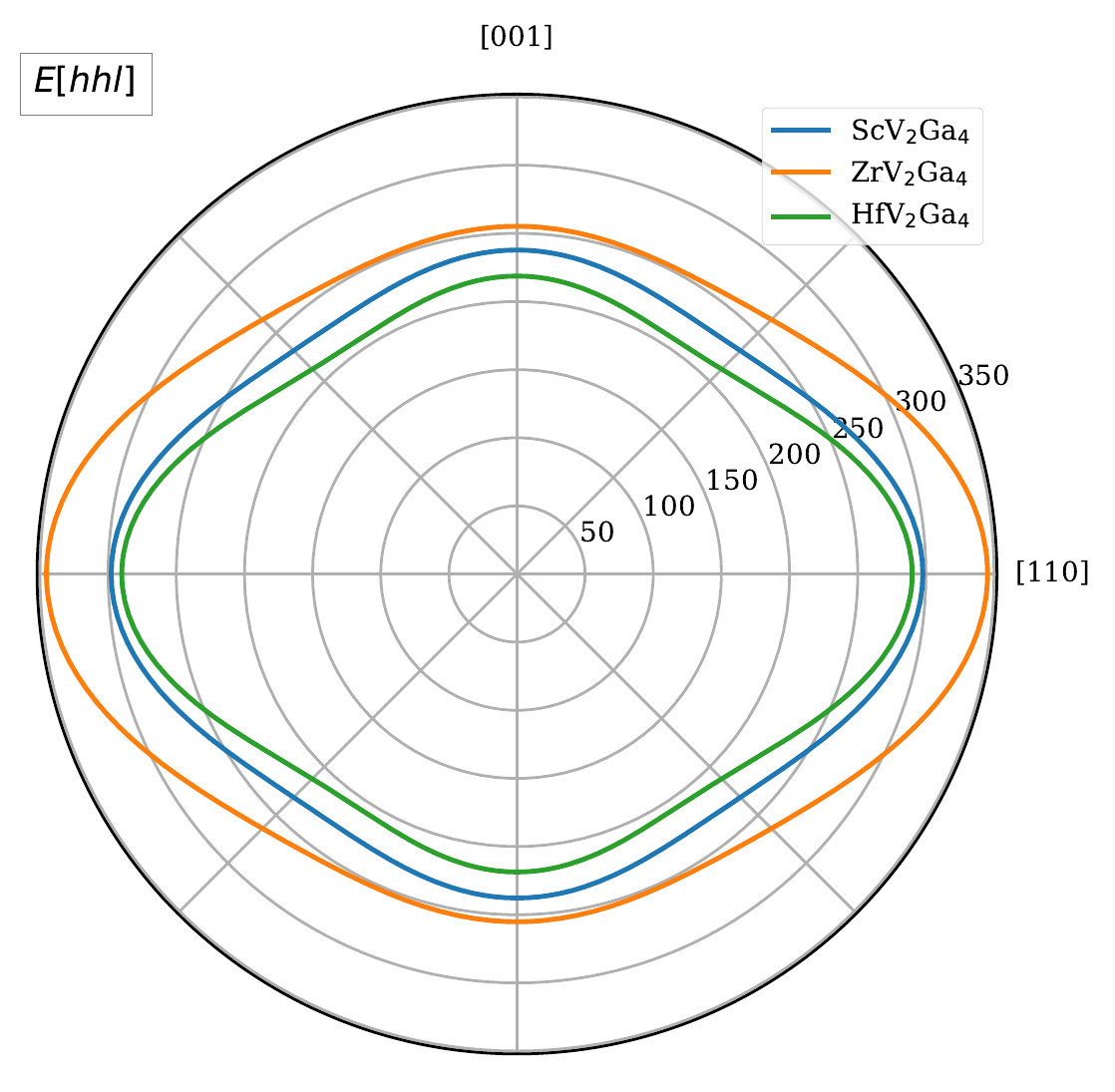}} 
	\caption{(a-c) Young's modulus $E$ (in GPa) for directions (a) $[0kl]$ in the (100) plane, (b) $[hk0]$ in the (001) plane, and (c) $[h\bar{h}l]$ in the (110) plane for MV$_2$Ga$_4$ compounds (M = Hf, Sc, Zr).}
	\label{fig:planar}
\end{figure*}

To provide an even more straightforward picture about the directional dependence of Young's modulus, a few planar projections are illustrated in Figures \ref{fig:planar}(a-c) for directions in the (100), (001) and (110) crystallographic planes. The anisotropy coming from the extended linear vanadium chains is well visualized in the (100) plane, showing a larger resistence to unixial stress along the [001] direction than the [010] direction. It's also clear that ZrV$_2$Ga$_4$ presents the largest values for Young's modulus, independent of the crystallographic orientation, and  also the smallest anisotropy regime among all considered structures. The directional dependence for ScV$_2$Ga$_4$ and HfV$_2$Ga$_4$ is quite similar in the (001) plane, but the effect of the extended linear chains is felt in the (100) and (110) planes. As observed in a previous work \cite{ferreira2018}, the valence and conduction electronic density along the vanadium chains is larger for ScV$_2$Ga$_4$ than HfV$_2$Ga$_4$ and, therefore, the V--V bonds are stronger in the former compound, explaining its higher Young's modulus along the [001] direction.

%%%%%%%%%%%%%%%%%%%%%%%%%%%%%%%%%%%%%%%%%%%%
% Poisson

It is harder to analyze the anisotropy in Poisson's ratio, since it depends not only on the elongation direction, but also on which normal direction one is interested in. Therefore, 3D plots such as those shown in Figures \ref{fig:mechanical_anisotropy}(a-f) are not possible. On the other hand, we can analyse the effect of longitudinal elongation along a few selected directions in the form of polar plots. % as those shown in Figures \ref{fig:poisson}(a-c).
In fourth-rank tensor notation, Poisson's ratio is given by \cite{Hayes1998}
\begin{equation}
 -\frac{\nu(\vec l,\, \vec n)}{E(\vec l)} = s_{ijkm} n_i n_j l_k l_m
\end{equation}
where the normal direction $\vec n$ can be defined by the angle $\psi$ in a plane perpendicular to $\vec l$, as shown in Figure \ref{fig:angles}. Algebraically we can write $\vec n$ as 
\begin{equation}
 \vec n = 
 \begin{pmatrix}
  n_1 \\ n_2 \\ n_3
 \end{pmatrix}
=
\begin{pmatrix}
  -\cos \varphi \cos \theta \sin \psi - \sin \varphi \cos \psi \\
    -\sin \varphi \cos \theta \sin \psi + \cos \varphi \cos \psi \\
      \sin \theta \sin \psi 
 \end{pmatrix}
\end{equation}
%where the angle $\psi$ is defined in Figure \ref{fig:angles}.
%
 In Voigt's notation for body-centered tetragonal structures, Poisson's ratio for the normal direction $\vec n$, when the elongation happens along direction $\vec l$, can be written as \cite{Goldstein2015}:
\begin{align}
 -\frac{\nu(\vec l,\, \vec n)}{E(\vec l)} = s_{11} (l_1^2 n_1^2 + l_2^2 n_2^2) + s_{12} (l_1^2 n_2^2 + l_2^2 n_1^2) + \nonumber \\
  +s_{13} (l_3^2 + n_3^2 - 2 l_3^2 n_3^2) +  s_{33} l_3^2 n_3^2 + \nonumber \\
  +s_{44} (l_1 n_1 +l_2 n_2) l_3 n_3 +s_{66} l_1 l_2 n_1 n_2
  \label{eq:poisson-tetra}
\end{align}

\begin{figure*}[t]
	  \centering
		\subfloat[][]{\includegraphics[width=.33\textwidth]{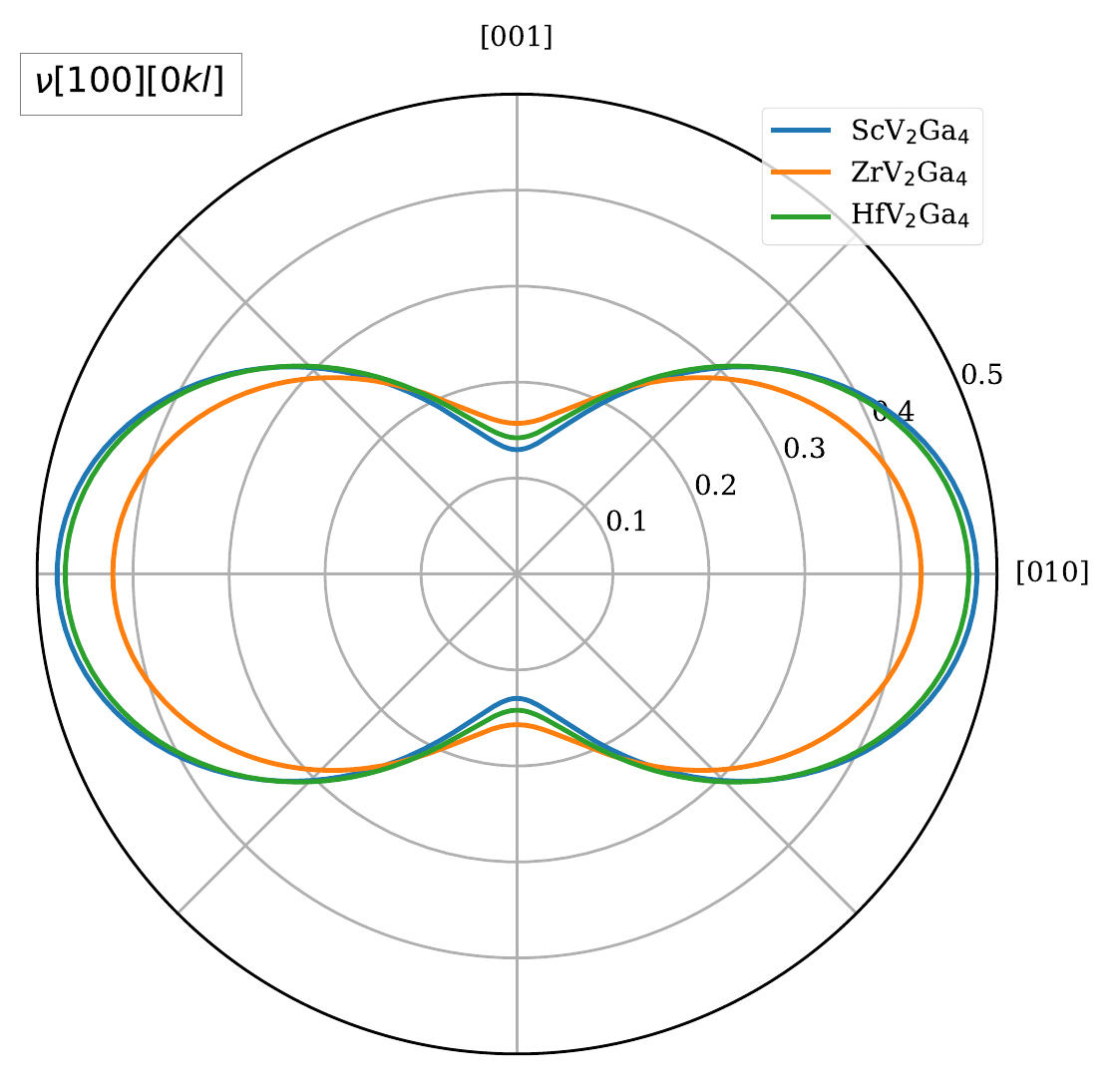}}
		\subfloat[][]{\includegraphics[width=.33\textwidth]{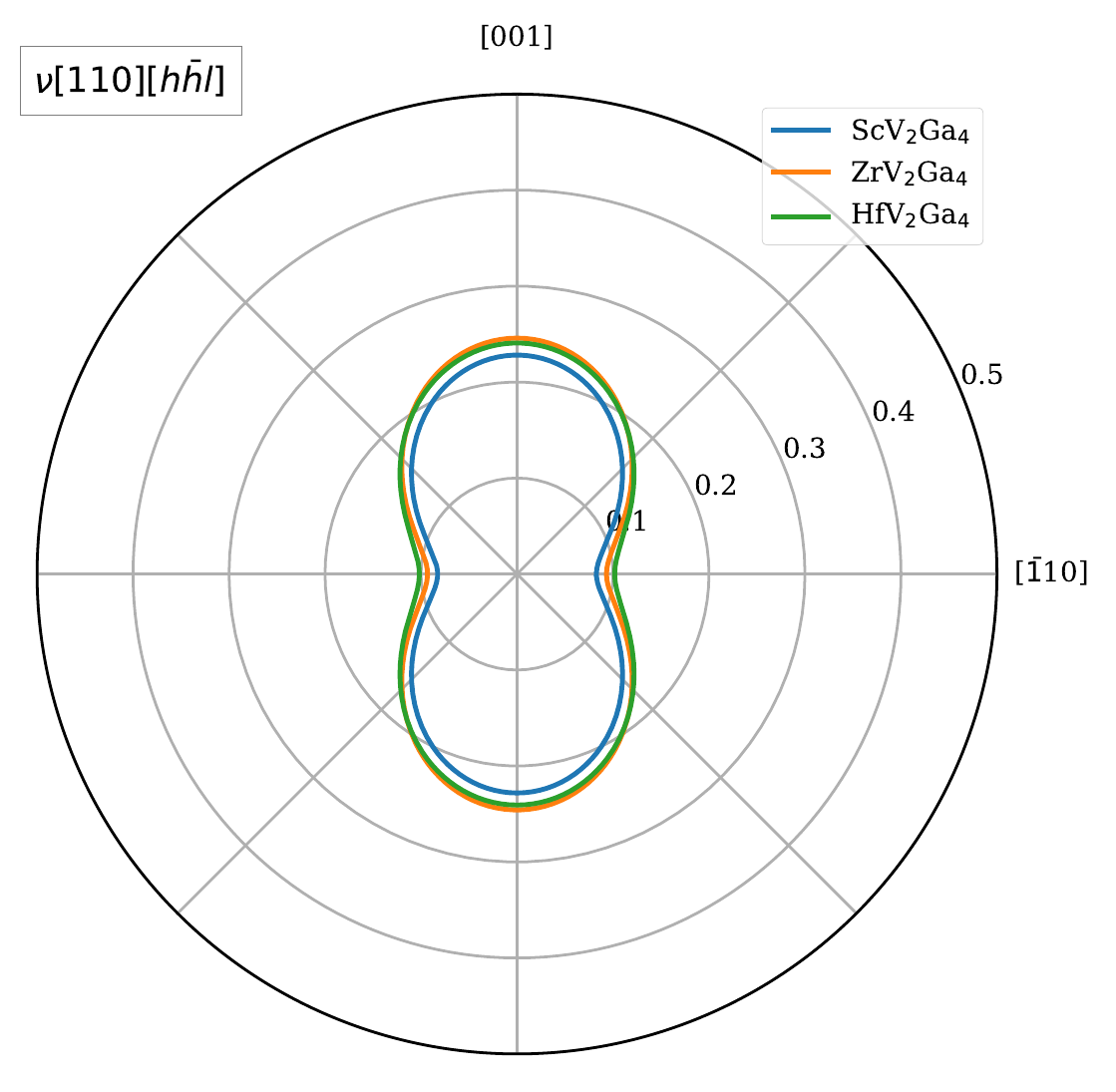}}
		\subfloat[][]{\includegraphics[width=.33\textwidth]{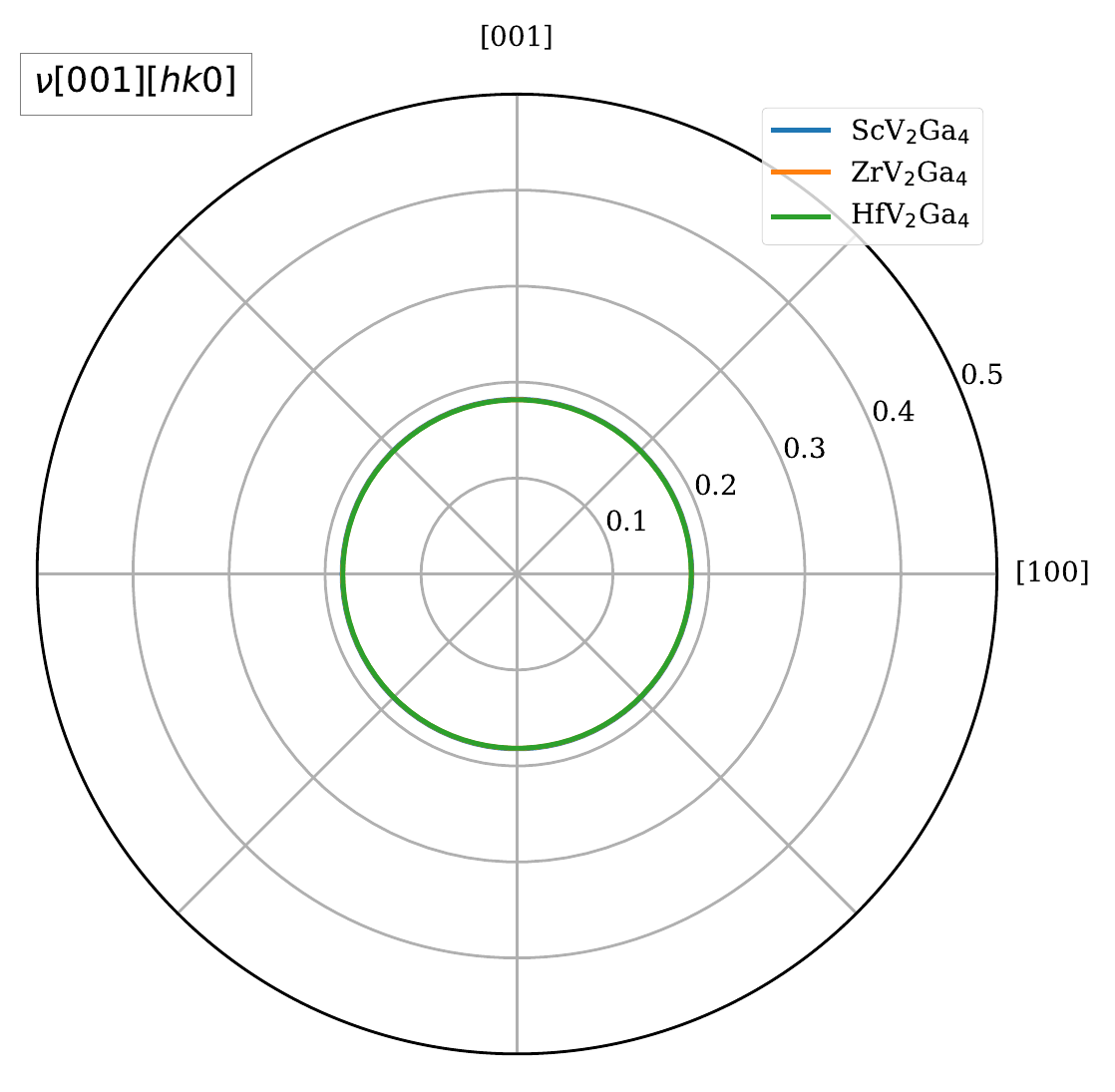}}
		\caption{Poisson ratio for MV$_2$Ga$_4$ compounds (M = Hf, Sc, Zr) measured in normal directions to the longitudinal extension along the (a) [100], (b) [110] and (c) [001] directions.}
		\label{fig:poisson}
\end{figure*}

Using this strategy, Poisson's ratios resulting from three different elongation directions are shown in Figure \ref{fig:poisson}(a-c). 
In Figure \ref{fig:poisson}(a), the elongation is along the [100] direction, and the Poisson's ratio is measured in the $[0kl]$ family of directions perpendicular to it. It is readily seen that the Poisson's ratio is quite large (close to 0.5) in the [010] direction. It should be kept in mind that, although the Poisson's ratio is limited to the interval $-1 \le \nu \le 0.5$ for isotropic materials, there is no such restriction when we consider deformation of single crystals \cite{Gunton1972}. Still, a value close to 0.5 indicates that areas in (001) planes are only weakly affected by elongation along the [100] direction. The behavior is drastically different in the [001] crystallographic orientation, where $\nu$ is quite small, around 0.15 for all compounds. This indicates that the dimensions along the tetragonal fourfold symmetry axes (i.e., the $c$-axis) are only slightly affected by elongation in the [100] direction. We readily see the effect of the extended linear chains as the explanation for this characteristic, since a contraction in the $c$-direction would mean a shortening of the directional V--V bonds, enormously affecting the energy density of the deformed structure. 
It is important to notice that this contraction in the $c$-direction, being the effect of an elongation along [100], happens without stress in the contraction direction. This is  different from the case shown in Figures \ref{fig:mechanical_anisotropy}(a-c), that come from an applied isostatic pressure, in which the [001] direction is the most affected.
Moreover, we see that ZrV$_2$Ga$_4$ has the lowest Poisson's ratio in the [010] direction, while ScV$_2$Ga$_4$ has the lowest value in the [001] direction. Since, as we mentioned previously, ScV$_2$Ga$_4$ has the highest electronic density along the extended linear V chains, is expected that it will be the most resistent against V--V bond length alterations.

Figure \ref{fig:poisson}(b) shows Poisson's ratio values when the elongation  happens in the [110] direction, that is, in directions normal to the highest Young's modulus directions in the crystal. In this case, all three componds behave in a similar manner, with low Poisson's ratio values, in the range 0.09--0.22. The values for the [001] direction are similar to those in Figure \ref{fig:poisson}(a), for the reasons explained in the preceeding paragraph. In the [$\bar{1}$10] crystallographic orientation, the values are even smaller, a consequence of the high resistence along this direction. 

Finally, the Poisson's ratio for elongation along the [001] direction is shown in Figure \ref{fig:poisson}(c). In this case, the tetragonal symmetry imposes transverse isotropy in  Poisson's ratio for directions in the (001) plane \cite{ballato96}, in which case Eq. (\ref{eq:poisson-tetra}) reduces to 
\begin{equation}
 \nu[001][hk0] = -\frac{s_{13}}{s_{33}}
\end{equation}
resulting in the circles shown in Figure \ref{fig:poisson}(c). In the scale of the figure, it is impossible to separate the curves for the three compounds, but they are slightly different: 0.1812 for ScV$_2$Ga$_4$, 0.1816 for ZrV$_2$Ga$_4$, and 0.1815 for HfV$_2$Ga$_4$. This apparent coincidence is again explained by the presence of extended linear V chains: the V--V and V--Ga bonds are those responsible for most of the contraction (or resistence against it) along directions perpendicular to the $c$-axis, {regardless of the atom occupying the M sites}. It should be noticed that this equivalence of Poisson's ratio for elongation along the fourfold symmetry axes is not found among other tetragonal systems \cite{feng2013}, which, one more time, points to the peculiarity of extended linear chain compounds.

%%%%%%%%%%%%%%%%%%%%%%%%%%%%%%%%%%%%%%%
% Shear modulus

\begin{figure*}[t]
	\centering
	\subfloat[][]{\includegraphics[width=.33\textwidth]{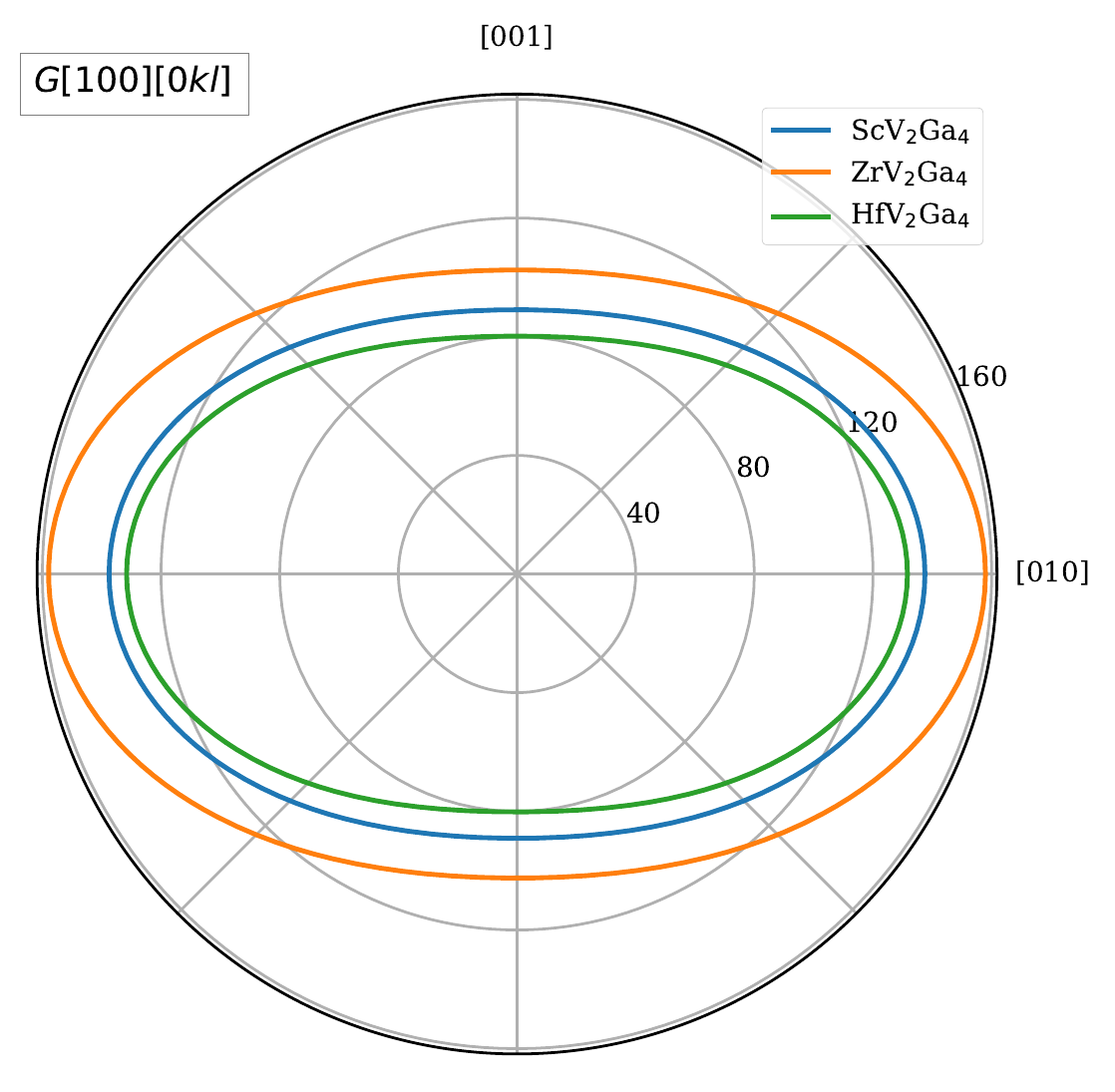}}
	\subfloat[][]{\includegraphics[width=.33\textwidth]{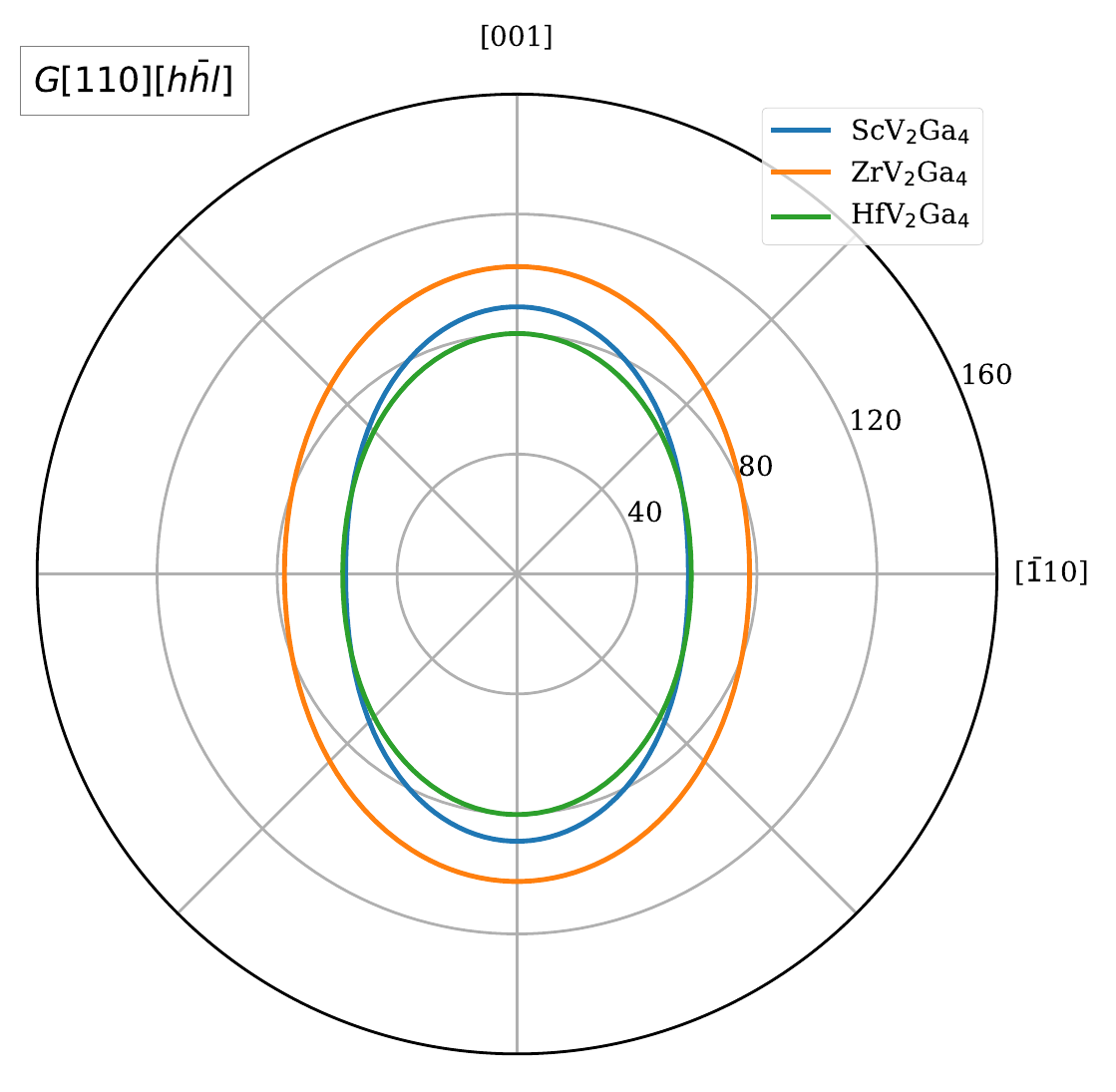}}
	\subfloat[][]{\includegraphics[width=.33\textwidth]{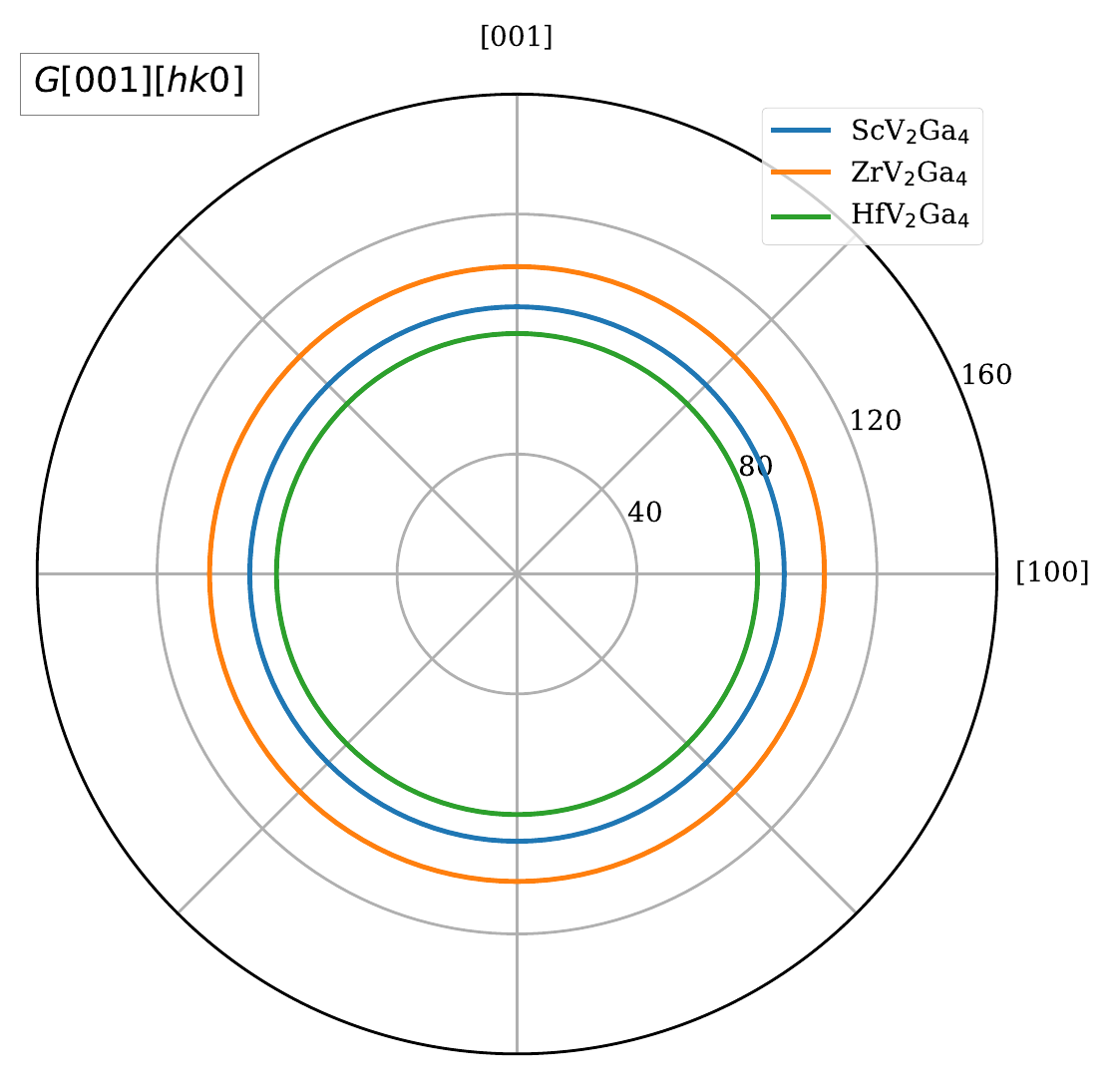}} 
	\caption{Shear modulus (in GPa) for MV$_2$Ga$_4$ compounds (M = Hf, Sc, Zr) measured in normal directions to the longitudinal (a) [100], (b) [110] and (c) [001] directions.}
	\label{fig:shear}
\end{figure*}

We can adopt a similar procedure to investigate the anisotropy in the shear modulus, which, in fourth-rank tensorial notation, is given by \cite{Hayes1998}
\begin{equation}
 G^{-1}(\vec l,\, \vec n) = 4 s_{ijkm} l_i n_j l_k n_m
\end{equation}
For a tetragonal body-centered structure using Voigt's matrix notation, the last equation amounts to
\begin{align}
 G^{-1}(\vec l,\, \vec n) = 4 s_{11} (l_1^2 n_1^2 + l_2^2  n_2^2) + \nonumber \\
        + 4 s_{33} l_3^2 n_3^2 + 8 s_{12} l_1 l_2 n_1 n_2 + \nonumber \\
        + 8 s_{13}  (l_1 n_1 + l_2 n_2) l_3 n_3 + s_{66} (l_1 n_2 + l_2 n_1)^2 + \nonumber \\
        + s_{44} \left[ (n_1^2 + n_2^2) l_3^2 + (l_1^2 + l_2^2) n_3^2 + 2  (l_1 n_1 + l_2 n_2)  l_3 n_3 \right].
        \label{eq:G-tetra}
\end{align}

Polar representations for selected stress states are shown in Figures \ref{fig:shear}(a-c). Shear modulus in the (100) plane, i.e., for $\vec n = [0kl]$ and $\vec l$ parallel to the $x$-axis, is shown in Figure \ref{fig:shear}(a). It is seen that a larger shear modulus is found in the [010] direction, that is, for shear in the $y$-direction. All three compounds exhibit a lower resistance against shear when $\vec n = [001]$, a result that comes from the fact that $s_{66}<s_{44}$ for the three compounds. With applied uniaxial stress in the [110] direction, as shown in Figure \ref{fig:shear}(b), the situation is reversed: higher shear modules for $\vec n=[001]$ (which is the same as in Fig. \ref{fig:shear}a) than for a direction normal to it, $[\bar{1}10]$. The lower shear modules in this case can be accounted for by the higher atomic density in the (110) planes, favouring slip.
Finally, for $\vec l=[001]$, that is, along the fourfold symmetry axes, the transverse isotropy of tetragonal lattices reduces  Eq. (\ref{eq:G-tetra}) simply to 
\begin{equation}
 G[001][hk0] = \frac{1}{s_{44}} = c_{44}
\end{equation}
resulting in the concentric circles shown in Figure \ref{fig:shear}(c). Therefore, as mentioned at the end of Sec. \ref{sec:stiff}, it is natural that ZrV$_2$Ga$_4$, possessing the higher $c_{44}$ value, is the compound with the higher shear modulus on the (001) plane.

\subsection{Sound velocity and Debye temperature}

The anisotropy can also be evaluated by means of the sound velocities in an elastic medium. The pure longitudinal and transverse modes for elastic waves in single crystals are essentially dictated by its symmetry \cite{brugger1965}. For instance, a tetragonal symmetry implies that the principal directions along which elastic waves can propagate in pure longitudinal and transverse modes are [001], [100] and [110]. In all other crystallographic orientations the propagating waves are either quasi-transverse or quasi-longitudinal. In the propagating directions for a tetragonal lattice the sound velocities can be written simply as:
\begin{itemize}	
\item {[100] direction:}
\begin{equation}
 \upsilon_l = \sqrt{c_{11}/\rho}; \quad \upsilon_{t+} = \sqrt{c_{44}/\rho}; \quad \upsilon_{t-} = \sqrt{c_{66}/\rho},  \\
\end{equation}
\item  {[001] direction:} 
\begin{equation}
\upsilon_l = \sqrt{c_{33}/\rho}; \quad \upsilon_{t+} = \sqrt{c_{66}/\rho}; \quad \upsilon_{t-} = \sqrt{c_{66}/\rho},  
\end{equation}
\item  {[110] direction:} 
\begin{align}
\upsilon_l = \sqrt{(c_{11} + c_{12} + 2c_{66})/2\rho}; \quad \upsilon_{t+} = \sqrt{c_{44}/\rho};  \nonumber \\ 
\upsilon_{t-} = \sqrt{(c_{11} - c_{12})/2\rho}, 
\end{align}
\end{itemize}
where $\rho$ is the mass density, $\upsilon_l$ is the longitudinal sound velocity, and $\upsilon_{t+}$ and $\upsilon_{t-}$ are the first and second transverse acoustic modes, respectively. 

\begin{table*}
	\caption{Anisotropic sound velocities (in m/s) along the three principal axes of MV$_2$Ga$_4$ (M = Sc, Zr, Hf). }
	\label{tab:sound}
	\centering
	\small
	\begin{tabular}{cccccccccc}
		\hline
		Direction & \multicolumn{3}{c}{$\left[100\right]$} & \multicolumn{3}{c}{$\left[001\right]$} & \multicolumn{3}{c}{$\left[110\right]$} \\
		&  $\upsilon_l$ & $\upsilon_{t+}$ & $\upsilon_{t-}$ & $\upsilon_l$ & $\upsilon_{t+}$ & $\upsilon_{t-}$ & $\upsilon_l$ & $\upsilon_{t+}$ & $\upsilon_{t-}$\\
		\hline		
		ScV$_2$Ga$_4$ & 5951.38 & 3659.90 & 4546.54 & 6268.77 & 4546.54 & 4546.54 & 6891.90 & 3659.90 & 2931.15\\[2mm]
		ZrV$_2$Ga$_4$ & 6206.89 & 3698.39 & 4590.31 & 6145.31 & 4590.31 & 4590.31 & 7017.68 & 3698.39 & 3216.93\\[2mm]
		HfV$_2$Ga$_4$ & 5262.57 & 3061.98 & 3922.32 & 5321.11 & 3922.32 & 3922.32 & 6023.88 & 3061.98 & 2606.17\\[2mm]
		\hline
	\end{tabular}
\end{table*}

Considering the expressions above, it is easy to verify that the sound velocities present similar elastic anisotropy as the second-order elastic stiffness constants, and smaller densities with higher elastic properties result in large propagating modes inside the crystal. The complete set of longitudinal and transverse acoustic modes along the principal axes was calculated and listed in Table \ref{tab:sound}. The elastic sound waves travel faster along the [110] longitudinal direction in ScV$_2$Ga$_4$, ZrV$_2$Ga$_4$ and HfV$_2$Ga$_4$ as an immediate consequence of the higher mechanical properties in that direction. In fact, ZrV$_2$Ga$_4$ is responsible for the strongest pure modes in all principal crystallographic orientations, except along the extended vanadium chains. Indeed, the large longitudinal sound velocity in [001] happens for ScV$_2$Ga$_4$. This is an interesting find: since ScV$_2$Ga$_4$ has the largest elastic anisotropy as a result of high populated V-chains along the $c$-direction, its atoms effectively interact with each other and consequently vibrate at higher frequencies like a tight rope, even ZrV$_2$Ga$_4$ presenting higher values for linear compressibility and Young's modulus along that orientation. In a similar way, the transverse modes for the three compounds are almost the same. Thus we can clearly observe the influence of the extended linear chain substructures in single crystals regarding its mechanical response. Despite the fact that HfV$_2$Ga$_4$ shows a substantial elastic anisotropy as seen in the spherical and polar plots, as well as in the A$^U$ universal index, its heavier Hf atoms cause an appreciable impact on the compound density, leading to lower longitudinal and transverse modes for elastic waves. Therefore, the sound velocities obtained at different crystallographic orientations provide valuable information that tie together all conclusions we made about elastic anisotropy of the MV$_2$Ga$_4$ compounds.

\begin{table}
	\caption{Longitudinal ($\upsilon_l$),transverse ($\upsilon_t$), and  mean ($\upsilon_m$) sound velocities (in m/s), and Debye temperature $\Theta_D$ (in K) for the MV$_2$Ga$_4$ (M = Sc, Zr, Hf) compounds.}
	\label{tab:debye}
	% 	\footnotesize
	\centering
	\begin{tabular}{ccccc}
		\hline
		& $\upsilon_l$ & $\upsilon_t$  & $\upsilon_m$ & $\Theta_D$ \\
		\hline
		ScV$_2$Ga$_4$ & 6221.15 & 3683.89 & 4080.42 & 490.21 \\[2mm]
		ZrV$_2$Ga$_4$ & 6373.94 & 3818.91 & 4225.23 & 510.33 \\[2mm]
		HfV$_2$Ga$_4$ & 5381.91 & 3147.86 & 3490.67 & 416.39 \\				
		\hline
	\end{tabular}
\end{table}

Sound velocities are essential to characterize the elastic and thermal response of a given solid mostly because they are easily obtained by experiments. On the same footing, the Debye temperature considers the highest normal mode of quantized vibration in a crystal and can be easily accessed by means of the average sound velocities, as expressed in the following equation: \cite{timoshenko2008},
\begin{align}
	\Theta_D = \dfrac{\hbar}{k}\left(\frac{6\pi^2}{V_{at}}\right)^{\frac{1}{3}}\upsilon_m,
\end{align} 
where $\hbar=1.054 \times 10^{-34}\,$J\,s and $k=1.381\times10^{-23}\,$J\,K$^{-1}$ are the reduced Planck constant ($h/2\pi$) and Boltzmann constant, respectively; $V_{at}$ is the volume of the unit cell divided by the total number of atoms in the unit formula; and $\upsilon_m$ is the effective average sound velocity in the material, given by
\begin{align}
	\dfrac{3}{\upsilon_m^3} = \dfrac{1}{\upsilon_l^3} + \dfrac{2}{\upsilon_t^3}.
\end{align}  
In the expression above, $\upsilon_l$ and $\upsilon_t$ are the longitudinal and transverse propagating modes, in which
\begin{equation}
	\upsilon_l = \sqrt{\dfrac{\lambda + 2\mu}{\rho}}
\end{equation}
and
\begin{equation}
	\upsilon_t = \sqrt{\dfrac{\mu}{\rho}}, 
\end{equation}
with $\lambda$ and $\mu$ being the Lamé parameters, determined from Poisson's ratio and bulk modulus:
\begin{equation}
	\lambda = \dfrac{3\nu}{1+\nu}B 
\end{equation}
\begin{equation}
	\mu = \dfrac{3}{2}\dfrac{1-2\nu}{1+\nu}B.
\end{equation}
Applying this procedure, we have calculated the effective longitudinal, transverse and average sound velocities, as well as the related Debye temperature for each structure, which are presented in Table \ref{tab:debye}. Consistently with our previously analysis, ZrV$_2$Ga$_4$ assumes the highest modes, since its higher bulk modulus values lead to faster sound velocities. On the other hand, the ScV$_2$Ga$_4$ compound also reveals substancially large elastic wave velocities since, as discussed earlier, its high electronic density confers very effective modes.

The calculated Debye temperatures also support our previous results, since this quantity reflects to a certain extent the strength of the interatomic bonding. The calculated 416.39\,K value for the Debye temperature of HfV$_2$Ga$_4$ is in excellent agreement with the 418.97\,K value obtained in a recent experimental investigation \cite{Santos2018}. We can also point out that ZrV$_2$Ga$_4$ and ScV$_2$Ga$_4$ should have better heat transfer properties than HfV$_2$Ga$_4$ since the thermal conductivity response is directly related to the average sound velocity \cite{Cahill1988}.

At this point of the discussion, our attention is driven to the possible consequences that the above results will have on the superconducting properties. It is well-known that the critical temperature in which a electron-phonon superconductor undergoes a second-order phase transition from its normal state to a superconducting state, as calculated using McMillan's formula, basically depends on the density of states at the Fermi level $N(E_F)$, the strength of the the electron-phonon coupling $\lambda$, the Debye temperature $\Theta_D$, and the electron-electron interaction constant (or Coulomb pseudopotential) $\mu^{*}$ \cite{mcmillan1968, ferreira2018}. {The higher the first three parameters, the higher the critical superconducting temperature will be, while increasing the last parameter increases the electron scattering and, therefore, tends to reduce $T_c$.} Thus, we argue that the ZrV$_2$Ga$_4$ compound, once experimentally proved to be a supercondutor, could present a $T_c$ compatible with ScV$_2$Ga$_4$, or even higher, as a consequence of the calculated value of 510.33\,K for the Debye temperature and the high elastic wave velocities, which will result in a more effective electron-phonon interaction in the lattice.  In the same way, our evaluated sound velocity values reinforce the previous theoretical prediction that ScV$_2$Ga$_4$ presumably consists of an electron-phonon superconductor with a considerable higher $T_c$ than HfV$_2$Ga$_4$ \cite{ferreira2018}.

\subsection{Debye-Gr\"{u}neisen approximation}

In order to obtain a few thermal properties of the MV$_2$Ga$_4$ (M = Sc, Zr, Hf) compounds, we have applied the quasi-harm\-onic Debye-Gr\"uneisen approximation \cite{Moruzzi1988, Duancheng2015, Liu2015}, which provides a strategy to calculate the temperature-dependent heat capacity, and, ultimately, the Gibbs (free) energy, using only 0\,K data. {It should be noticed that this approach is only qualitatively correct, and other, more sophisticated and precise approaches are found in the literature \cite{Zhang2018, Zhang2017, Gupta2017, Glensk2015}. The Debye Gr\"{u}neisen method, on the other hand, is sufficient for our present purposes, as we are interested in the low temperature range.} 

\begin{figure*}[t]
	\centering
	\subfloat[][Bulk modulus]{\includegraphics[width=.5\textwidth]{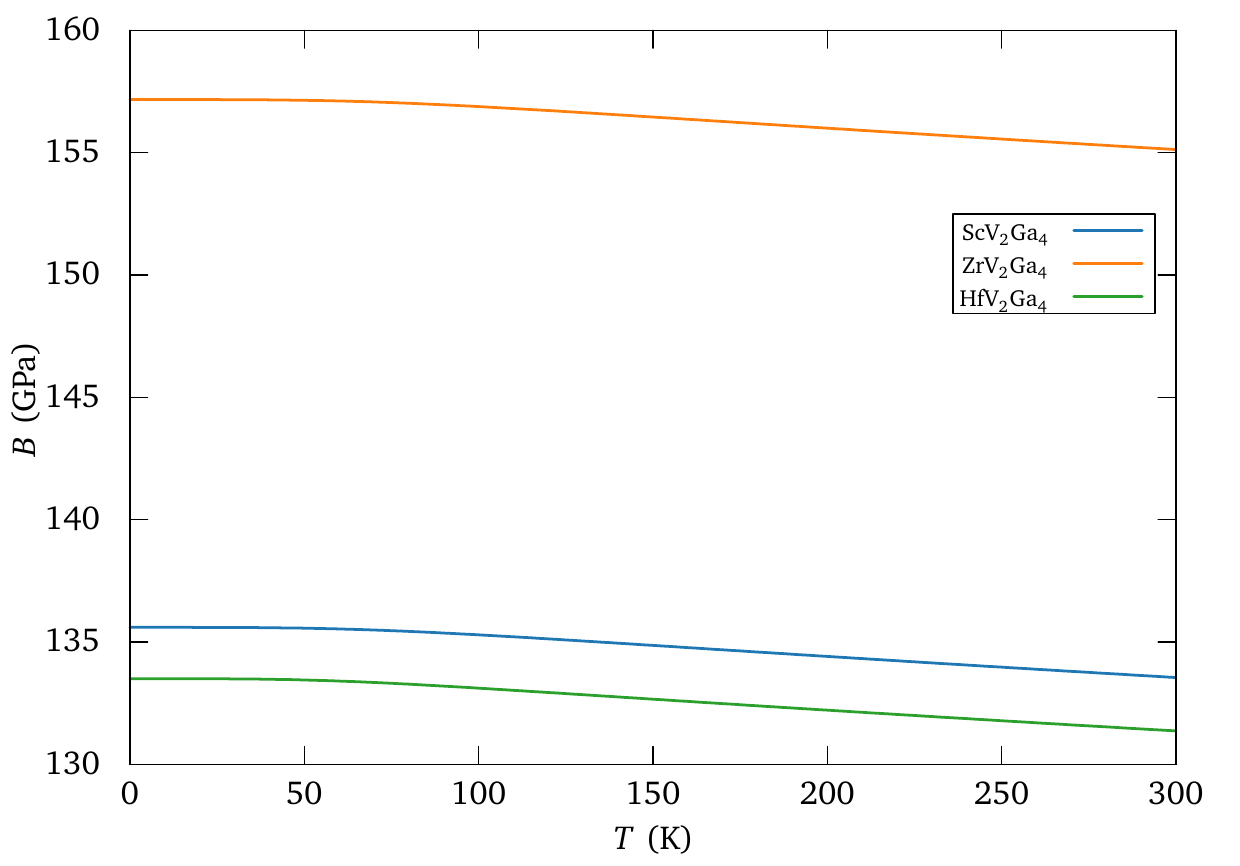}}
	\subfloat[][Young's modulus]{\includegraphics[width=.5\textwidth]{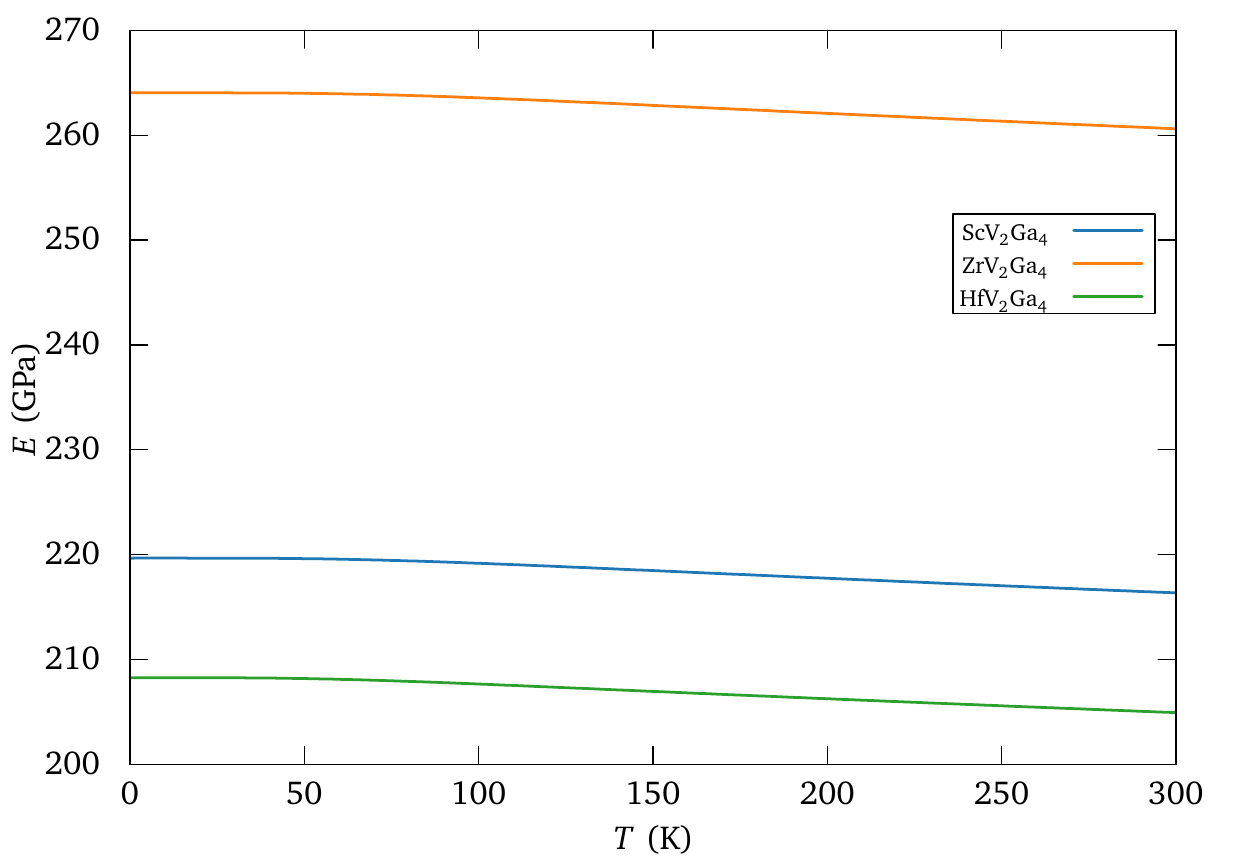}}\\
	\subfloat[][Thermal expansion]{\includegraphics[width=.5\textwidth]{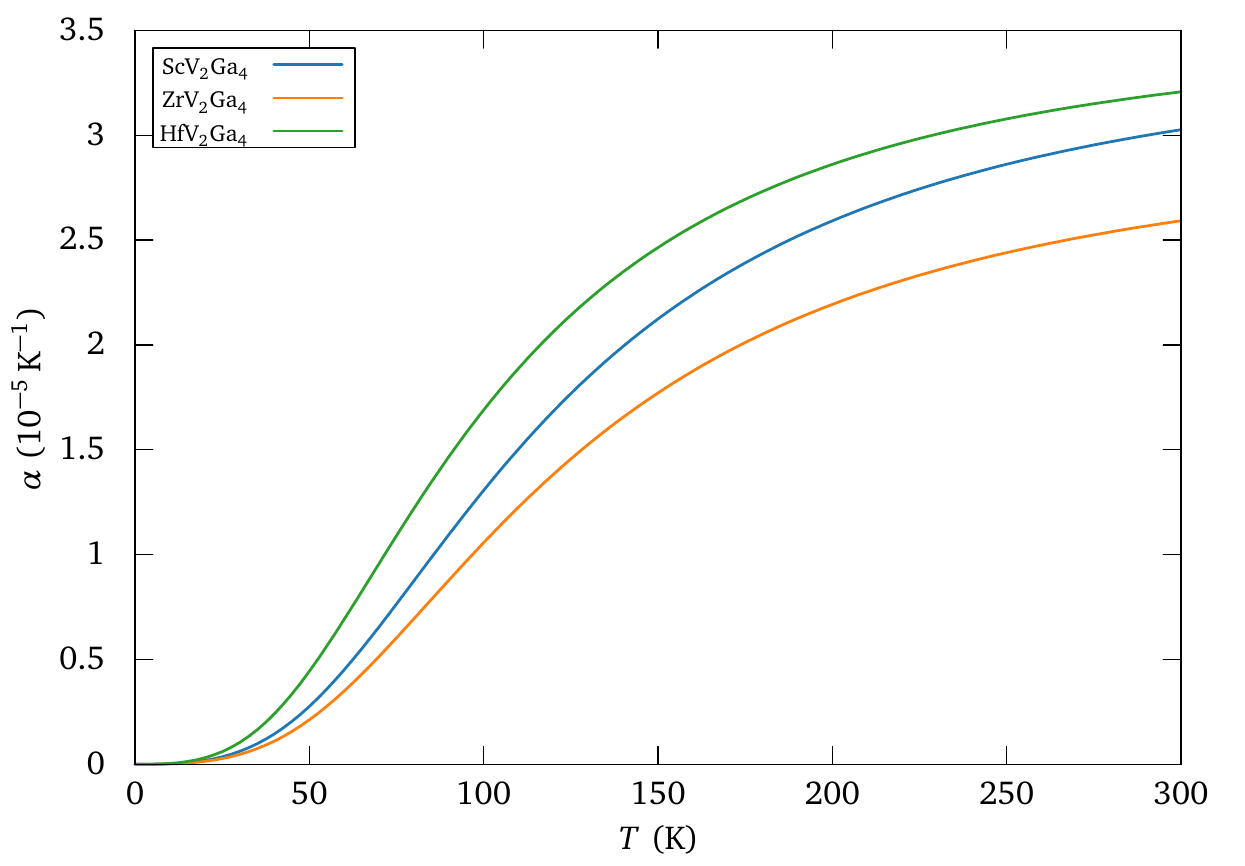}} 
	\subfloat[][Heat capacity at constant pressure]{\includegraphics[width=.5\textwidth]{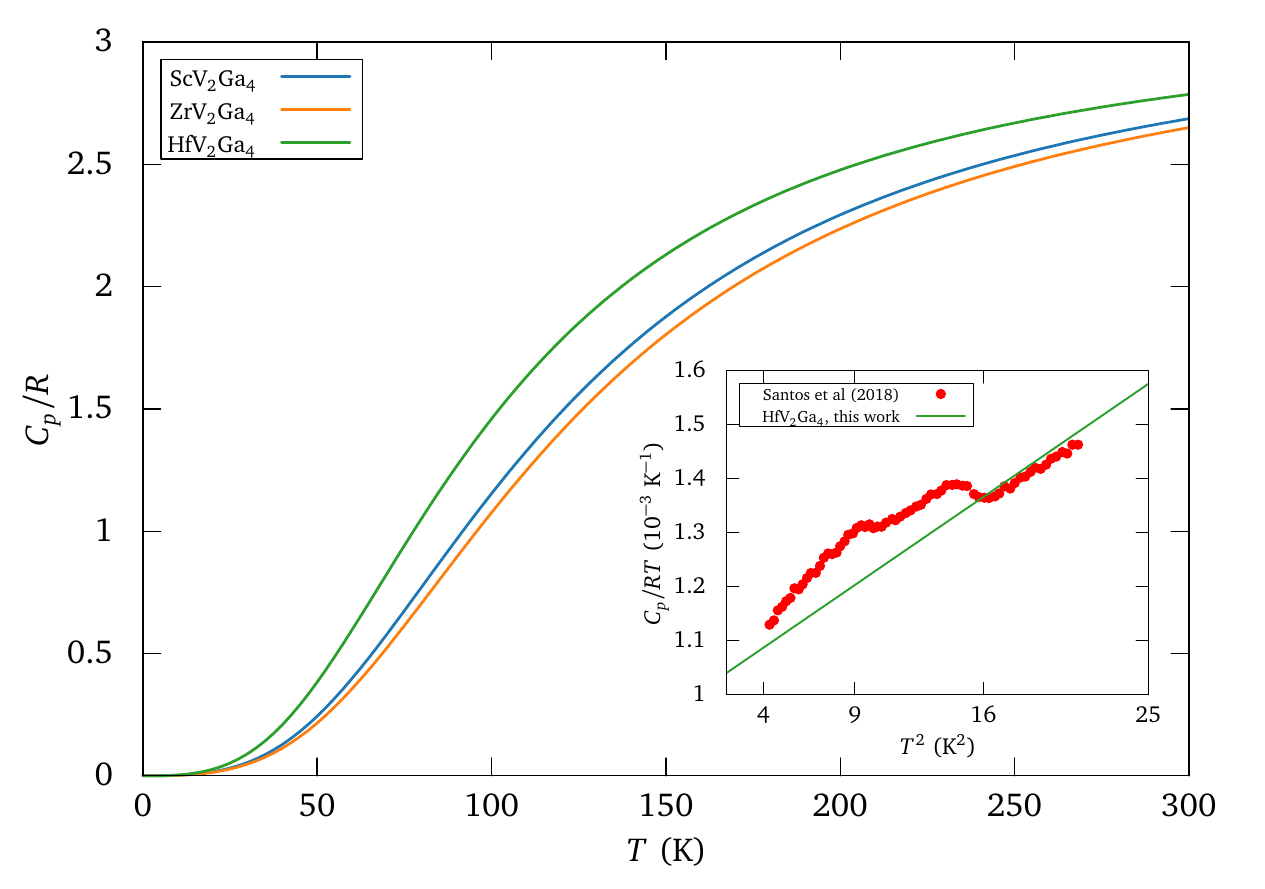}}
	\caption{Theoretical Debye-Gr\"{u}neisen quasi-harmonic calculation of (a) bulk modulus, (b) Young modulus, (c)  coefficient of volumetric thermal expansion, and (d) heat capacity at constant pressure for the MV$_2$Ga$_4$ (M=Hf, Sc, Zr) compounds. In (d), a comparison with experimental low temperature data \cite{Santos2018} for HfV$_2$Ga$_4$ is shown as an inset using a typical $C_p/T$ vs. $T^2$ plot ($R=8.31451\,$J/mol\,K is the universal gas constant).}
	\label{fig:thermal}
\end{figure*}

In a general way, the heat capacity of a stoichiometric, thermodynamically closed system at constant pressure can be written, {disregarding possible coupling effects \cite{Gupta2017}}, as
\begin{equation}
 C_p(p,\, T) = C_p^\text{el}(p,\,T) + C_p^\text{vib}(p,\,T) + C_p^\text{mag}(p,\,T)
\end{equation}
in which $T$ is the absolute temperature, $p$ is the applied external pressure and $C_p^\text{el}$, $C_p^\text{vib}$, and $C_p^\text{mag}$ are, respectively, the electronic, vibrational (phonon) and magnetic contributions. It should be kept in mind that this equation does not work for a general applied tension, only for an isostatic stress state corresponding to an external applied pressure $p$. Also, we are considering an isotropic elastic medium resulting from the VRH approximation. In this work there is no need to consider magnetic effects ($C_p^\text{mag}=0$), as there is no resulting spin polarization in the electronic structure of the compounds. The electronic contribution will be taken simply as $C_p^\text{el}(p,\,T)=\gamma_\text{el}=\text{const.}$, according to the Sommerfeld approximation. Finally, the vibrational heat capacity is given by
\begin{equation}
 C_p^\text{vib}(p,\,T) = C_V^\text{vib}(p,\,T)+ \alpha^2(p,\,T) B(p,\,T) T V(p,\,T)
\end{equation}
where $C_V^\text{vib}(p,\,T)$ is the vibrational heat capacity at constant volume, $\alpha(p,\,T)$ is the volumetric thermal expansion coefficient, $V(p,\,T)$ is the volume and $B(p,\,T)$ is the isothermal bulk modulus. According to the Debye-Gr\"uneisen approximation, 
\begin{equation}
 C_V^\text{vib}(p,\,T) = 9 N R \left( \frac{T}{\Theta_D} \right)^3 \int_0^{\Theta_D/T} \frac{x^4 e^x}{\left( e^x -1 \right)^2} dx
\end{equation}
in which $R=8.31451\,$J\,mol$^{-1}$\,K$^{-1}$ is the universal gas constant, $N$ is the number of atoms and the Debye temperature $\Theta_D$ is considered volume dependent:
\begin{equation}
 \Theta_D(V) = \Theta_D(V_0) \left( \frac{V_0}{V} \right)^\gamma
\end{equation}
where $\Theta_D(V_0)$ is the Debye temperature at 0\,K and zero external pressure, as seen in the preceeding section, and $\gamma$ is the Grüneisen parameter, that can be written as \cite{Duancheng2015}
\begin{equation}
 \gamma = g + \frac{1+B_p^\prime}{2},
\end{equation}
with $B_p^\prime$ being the 0\,K pressure derivative of the Bulk modulus at zero pressure, obtained from a fit to the ab-initio isotropic deformation data of an equation of state such as, for instance, the Birch-Murnaghan equation \cite{birch1947, poirier2000}. The coefficient $g$ can be fixed at $g=2/3$ if the aim is high-temperature data or $g=1$ for low temperatures \cite{Moruzzi1988}. The latter was the value adopted in this work.

Using the equations above, it is possible to calculate the temperature and pressure dependence of several quantities of interest. For instance, the volumetric thermal expansion is written as 
\begin{equation}
 \alpha(p,\,T) = \frac{\gamma C_V^\text{vib}(p,\,T)}{B(p,\,T) V(p,\,T)}
\end{equation}
with the Bulk modulus given by
\begin{equation}
 B^{-1}(p,\,T) = -\frac{1}{V(p,\,T)} \left(\frac{\partial V}{\partial p}\right)_T
\end{equation}
Finally, the Young's modulus is calculated using a result from the theory of elasticity of isotropic media:
\begin{equation}
 E(p,\,T) = \frac{3B(p,\,T)}{1-2\nu(p,\,T)}
\end{equation}
where $\nu(p,\,T)$ is the Poisson's ratio, which will be considered pressure- and temperature- independent in this work. Moreover, we will only be interested in the zero external pressure case and, therefore, the crystal structure will expand due to thermal effects only.

% Within the Debye-Gr\"uneisen approximation, the Gibbs energy is written as 

% The main results are presented in Fig.\ref{fig:thermal}(a-d). In a purely harmonic crystal as described by Debye model, the thermal expansion coefficient is identically zero. As a consequence, the specific heat at constant pressure and temperature assumes the same value without inharmonic effects, and hence, other correlated properties are independet of temperature, as bulk modulus and Debye temperature. In a first view, we can partially visualize the non-harmonic effects in the plot of $\Theta_D$ as function of temperature in Fig.\ref{fig:thermal}a. From Debye-Gruneisen model, it's possible identify that the Debye temperature is constant for temperatures lower than  $\approx$\,80\,K. With the increasing temperature, the highest phonon mode is decreased, as well as the interatomic bonding is weakened, as expected. This results are consistet with the fact that HfV$_2$Ga$_4$ has the largest thermal expansion coefficient and a sharp depence with increasing temperature when compared with ZrV$_2$Ga$_4$, which possess the smallest coefficient and a smoth variation, as showin Fig.\ref{fig:thermal}b 

The results of the calculations are shown in Figure \ref{fig:thermal}(a-d). {We limited the calculations to low temperatures since this is the region of interest and also because the method employed, as already discussed, is not precise enough outside this range.} It is seen, as expected, that the material softens with increasing temperature, as observed by the lowering of Bulk and Young's moduli in Figures \ref{fig:thermal}(a) and (b), respectively. Since the Debye temperature for all compounds are similar, the thermal softening ratio is approximately the same in all cases. 

The volumetric thermal expansion is shown in Figure \ref{fig:thermal}(c). {It is seen that ZrV$_{2}$Ga$_4$ is the compound presenting the lowest values, agreeing with the fact that it is the compound with the highest bulk modulus and, therefore, also the most resistant against variations in bond lengths.}
Finally, Figure \ref{fig:thermal}(d) shows the heat capacity at constant pressure as a function of temperature. Again, since the Debye temperatures are quite close to one another, the three curves are very similar. The inset in Figure \ref{fig:thermal}(d) shows a comparison to experimental data \cite{Santos2018} for HfV$_2$Ga$_4$ in a $C_p/T$ vs. $T^2$ plot, highlighting the $T^3$ dependence of the heat capacity at low temperatures. In order to obtain the calculated curve, besides the DFT calculations, we took the value $\gamma_\text{el}=8.263\,$mJ/mol\,K$^2$ for the Sommerfeld coefficient from the same experimental work \cite{Santos2018}. Unfortunately, there are no data in the literature for the other compounds. {The double jumps in the experimental curve, naturally, come from the second order transitions up to the critical temperature, around 4.1\,K, that cannot be reproduced whithin the scope of conventional DFT calculations.}
% The calculated depence with temperature of specific heat at constant volume and pressure is illustrated in Fig.\ref{fig:thermal}c and d, respectively. We can further verify the T$^3$ law. At high temperatures, $C_v$ reaches the Dulong-Petit limit, while $C_p$ increases monotonously by definition. 
It should be emphasized that our calculated $C_p$ curve for HfV$_2$Ga$_4$ shows {a good agreement with the available experimental data \cite{Santos2018} at low temperatures above $T_c$}, considering the intrinsic limitations contained in the approximations required by the numerical methods. 
%In this way, given its great relevance to describe the superconducting state, the calculated $C_p$ curves for ScV$_2$Ga$_4$ and ZrV$_2$Ga$_4$ can be of great importance in the context of future experimental or theoretical studies.

\section{Conclusions}

In this work we have performed ab-initio calculations, based on DFT-methods, of elastic and thermal properties of extended linear chain compounds MV$_2$Ga$_4$ (M = Sc, Zr, Hf). From the independent set of elastic stiffness and compliance constants, we have evaluated the mechanical properties of the polycristal\-line materials using the Voigt-Reuss-Hill approach. We found that ZrV$_2$Ga$_4$ has the largest values for bulk, shear and Young's modulus. Attached to this, the $B/G$ ratio indicates that all compounds have a brittle character, which is consistent with the obtained Poisson's ratio values. The universal elastic anisotropy index, as well as the directional dependence analysis of linear compressibility, Young's modulus, Poisson's ratio and shear modulus, all point out that ScV$_2$Ga$_4$ has the largest elastic anisotropy. We have shown that the extended linear V chains play a fundamental role in the mechanical properties, as observed in the Young's modulus, Poisson's ratio and shear modulus for different longitudinal and normal directions. For instance, the ZrV$_2$Ga$_4$ compound presents the strongest pure modes of elastic waves in all principal crystallographic orientations, except along the extended linear vanadium chains. In that direction, ScV$_2$Ga$_4$ exhibits the largest longitudinal sound velocity, a direct consequence of its highly populated V-chains.

From the calculated Debye temperatures and the high pure longitudinal and transverse modes for elastic wave velocities, which result in a more effective electron-phonon coupling in the structure, we also argue that ZrV$_2$Ga$_4$ could show a similar superconducting critical temperature, or even higher, than ScV$_2$Ga$_4$ (provided it is indeed a superconductor), a prediction that should be confirmed experimentally.

Finally, we believe that the approach and the results found in the present work will substantially advance our understanding of the properties in an emergent class of superconducting materials and extended linear chain compounds, forecasting and elucidating relevant parameters and motivating, as a guideline, future experimental investigations in similar or related systems.

\section*{Acknowledgements}

The research was carried out using high-performance computing resources made available by the Su\-per\-in\-ten\-dên\-cia de Tec\-no\-lo\-gia da In\-for\-ma\-ção (STI), University of São Paulo. This study was financed in part by the Coordenação de Aperfeiçoamento de Pessoal de Nível Superior - Brasil (CAPES) - Finance Code 001.
The authors also gratefully acknowledge the financial support of the Fun\-da\-ção de Am\-pa\-ro à Pes\-qui\-sa do Es\-ta\-do de São Pau\-lo (FAPESP, Proc. 2016/11565-7, 2016/11774-5 and 2017/11023-2), and the  Con\-se\-lho Na\-ci\-o\-nal de De\-sen\-vol\-vi\-men\-to Ci\-en\-tí\-fi\-co e Tec\-no\-ló\-gi\-co (Cnpq, Proc. 302149/2017-1).
%, and the Co\-or\-de\-na\-ção de A\-per\-fei\-ço\-a\-men\-to de Pes\-so\-al de Ní\-vel Su\-pe\-ri\-or (Capes).

\section*{References}

\begin{singlespace}

\end{singlespace}


\begin{thebibliography}{70}
	\providecommand{\natexlab}[1]{#1}
	\providecommand{\url}[1]{\texttt{#1}}
	\providecommand{\href}[2]{#2}
	\providecommand{\path}[1]{#1}
	\providecommand{\eprint}[1]{\href{http://arxiv.org/abs/#1}{\path{#1}}}
	\providecommand{\DOIprefix}{doi:}
	\providecommand{\ArXivprefix}{arXiv:}
	\providecommand{\URLprefix}{URL: }
	\providecommand{\Pubmedprefix}{pmid:}
	\providecommand{\doi}[1]{\href{http://dx.doi.org/#1}{\path{#1}}}
	\providecommand{\Pubmed}[1]{\href{pmid:#1}{\path{#1}}}
	\providecommand{\BIBand}{and}
	\providecommand{\bibinfo}[2]{#2}
	\ifx\xfnm\undefined \def\xfnm[#1]{\unskip,\space#1}\fi
	%Type = Article
	\bibitem[{Bardeen et~al.(1957)Bardeen, Cooper and Schrieffer}]{Bardeen1957}
	\bibinfo{author}{Bardeen\xfnm[ J.]}, \bibinfo{author}{Cooper\xfnm[ L.N.]},
	\bibinfo{author}{Schrieffer\xfnm[ J.R.]}.
	\newblock \bibinfo{title}{Theory of superconductivity}.
	\newblock \bibinfo{journal}{Phys Rev}
	\bibinfo{year}{1957};\bibinfo{volume}{108}:\bibinfo{pages}{1175}.
	%Type = Article
	\bibitem[{Tomita and Murakami(2003)}]{tomita2003}
	\bibinfo{author}{Tomita\xfnm[ M.]}, \bibinfo{author}{Murakami\xfnm[ M.]}.
	\newblock \bibinfo{title}{High-temperature superconductor bulk magnets that can
		trap magnetic fields of over 17 tesla at 29 {K}}.
	\newblock \bibinfo{journal}{Nature}
	\bibinfo{year}{2003};\bibinfo{volume}{421}(\bibinfo{number}{6922}):\bibinfo{pages}{517}.
	%Type = Article
	\bibitem[{Noe and Steurer(2007)}]{noe2007}
	\bibinfo{author}{Noe\xfnm[ M.]}, \bibinfo{author}{Steurer\xfnm[ M.]}.
	\newblock \bibinfo{title}{High-temperature superconductor fault current
		limiters: concepts, applications, and development status}.
	\newblock \bibinfo{journal}{Supercond Sci Technol}
	\bibinfo{year}{2007};\bibinfo{volume}{20}(\bibinfo{number}{3}):\bibinfo{pages}{R15}.
	%Type = Article
	\bibitem[{Tomsic et~al.(2007)Tomsic, Rindfleisch, Yue, McFadden, Phillips,
		Sumption et~al.}]{tomsic2007}
	\bibinfo{author}{Tomsic\xfnm[ M.]}, \bibinfo{author}{Rindfleisch\xfnm[ M.]},
	\bibinfo{author}{Yue\xfnm[ J.]}, \bibinfo{author}{McFadden\xfnm[ K.]},
	\bibinfo{author}{Phillips\xfnm[ J.]}, \bibinfo{author}{Sumption\xfnm[ M.D.]},
	et~al.
	\newblock \bibinfo{title}{Overview of {MgB2} superconductor applications}.
	\newblock \bibinfo{journal}{Int J Appl Ceram Technol}
	\bibinfo{year}{2007};\bibinfo{volume}{4}(\bibinfo{number}{3}):\bibinfo{pages}{250}.
	%Type = Article
	\bibitem[{Malozemoff et~al.(2008)Malozemoff, Fleshler, Rupich, Thieme, Li,
		Zhang et~al.}]{malozemoff2008}
	\bibinfo{author}{Malozemoff\xfnm[ A.P.]}, \bibinfo{author}{Fleshler\xfnm[ S.]},
	\bibinfo{author}{Rupich\xfnm[ M.]}, \bibinfo{author}{Thieme\xfnm[ C.]},
	\bibinfo{author}{Li\xfnm[ X.]}, \bibinfo{author}{Zhang\xfnm[ W.]}, et~al.
	\newblock \bibinfo{title}{Progress in high temperature superconductor coated
		conductors and their applications}.
	\newblock \bibinfo{journal}{Supercond Sci Technol}
	\bibinfo{year}{2008};\bibinfo{volume}{21}(\bibinfo{number}{3}):\bibinfo{pages}{034005}.
	%Type = Article
	\bibitem[{Putti et~al.(2010)Putti, Pallecchi, Bellingeri, Cimberle, Tropeano,
		Ferdeghini et~al.}]{putti2010}
	\bibinfo{author}{Putti\xfnm[ M.]}, \bibinfo{author}{Pallecchi\xfnm[ I.]},
	\bibinfo{author}{Bellingeri\xfnm[ E.]}, \bibinfo{author}{Cimberle\xfnm[
		M.R.]}, \bibinfo{author}{Tropeano\xfnm[ M.]},
	\bibinfo{author}{Ferdeghini\xfnm[ C.]}, et~al.
	\newblock \bibinfo{title}{New {Fe-based} superconductors: properties relevant
		for applications}.
	\newblock \bibinfo{journal}{Supercond Sci Technol}
	\bibinfo{year}{2010};\bibinfo{volume}{23}(\bibinfo{number}{3}):\bibinfo{pages}{034003}.
	%Type = Article
	\bibitem[{Werfel et~al.(2011)Werfel, Floegel-Delor, Rothfeld, Riedel, Goebel,
		Wippich et~al.}]{werfel2011}
	\bibinfo{author}{Werfel\xfnm[ F.]}, \bibinfo{author}{Floegel-Delor\xfnm[ U.]},
	\bibinfo{author}{Rothfeld\xfnm[ R.]}, \bibinfo{author}{Riedel\xfnm[ T.]},
	\bibinfo{author}{Goebel\xfnm[ B.]}, \bibinfo{author}{Wippich\xfnm[ D.]},
	et~al.
	\newblock \bibinfo{title}{Superconductor bearings, flywheels and
		transportation}.
	\newblock \bibinfo{journal}{Supercond Sci Technol}
	\bibinfo{year}{2011};\bibinfo{volume}{25}(\bibinfo{number}{1}):\bibinfo{pages}{014007}.
	%Type = Book
	\bibitem[{Schwartz(2013)}]{schwartz2013}
	\bibinfo{author}{Schwartz\xfnm[ B.]}.
	\newblock \bibinfo{title}{Superconductor applications: {SQUIDs} and machines};
	vol.~\bibinfo{volume}{21}.
	\newblock \bibinfo{publisher}{Springer Science \& Business Media};
	\bibinfo{year}{2013}.
	%Type = Article
	\bibitem[{Santos et~al.(2018)Santos, Correa, de~Lima, Cigarroa, da~Luz, Grant
		et~al.}]{Santos2018}
	\bibinfo{author}{Santos\xfnm[ F.B.]}, \bibinfo{author}{Correa\xfnm[ L.E.]},
	\bibinfo{author}{de~Lima\xfnm[ B.S.]}, \bibinfo{author}{Cigarroa\xfnm[
		O.V.]}, \bibinfo{author}{da~Luz\xfnm[ M.S.]}, \bibinfo{author}{Grant\xfnm[
		T.W.]}, et~al.
	\newblock \bibinfo{title}{Unusual superconducting behavior in {HfV$_2$Ga$_4$}}.
	\newblock \bibinfo{journal}{Phys Lett A}
	\bibinfo{year}{2018};\bibinfo{volume}{382}:\bibinfo{pages}{1065}.
	%Type = Article
	\bibitem[{Zehetmayer(2013)}]{Zehetmayer2013}
	\bibinfo{author}{Zehetmayer\xfnm[ M.]}.
	\newblock \bibinfo{title}{A review of two-band superconductivity: materials and
		effects on the thermodynamic and reversible mixed-state properties}.
	\newblock \bibinfo{journal}{Supercond Sci Technol}
	\bibinfo{year}{2013};\bibinfo{volume}{26}(\bibinfo{number}{4}):\bibinfo{pages}{043001}.
	%Type = Article
	\bibitem[{Ferreira et~al.(2018)Ferreira, Santos, Machado, Petrilli and
		Eleno}]{ferreira2018}
	\bibinfo{author}{Ferreira\xfnm[ P.P.]}, \bibinfo{author}{Santos\xfnm[ F.B.]},
	\bibinfo{author}{Machado\xfnm[ A.J.S.]}, \bibinfo{author}{Petrilli\xfnm[
		H.M.]}, \bibinfo{author}{Eleno\xfnm[ L.T.F.]}.
	\newblock \bibinfo{title}{Insights into the unconventional superconductivity in
		{HfV$_2$Ga$_4$} and {ScV$_2$Ga$_4$} from first-principles
		electronic-structure calculations}.
	\newblock \bibinfo{journal}{Phys Rev B}
	\bibinfo{year}{2018};\bibinfo{volume}{98}:\bibinfo{pages}{045126}.
	%Type = Book
	\bibitem[{Miller(2012)}]{miller2012}
	\bibinfo{author}{Miller\xfnm[ J.S.]}.
	\newblock \bibinfo{title}{Extended linear chain compounds};
	vol.~\bibinfo{volume}{3}.
	\newblock \bibinfo{publisher}{Springer Science \& Business Media};
	\bibinfo{year}{2012}.
	%Type = Article
	\bibitem[{Cutforth et~al.(1977)Cutforth, Datars, {Van Schyndel} and
		Gillespie}]{cutforth1977}
	\bibinfo{author}{Cutforth\xfnm[ B.]}, \bibinfo{author}{Datars\xfnm[ W.]},
	\bibinfo{author}{{Van Schyndel}\xfnm[ A.]}, \bibinfo{author}{Gillespie\xfnm[
		R.]}.
	\newblock \bibinfo{title}{Electrical conductivity of linear chain mercury
		compounds}.
	\newblock \bibinfo{journal}{Solid State Commun}
	\bibinfo{year}{1977};\bibinfo{volume}{21}(\bibinfo{number}{4}):\bibinfo{pages}{37}.
	%Type = Article
	\bibitem[{Chiang et~al.(1977)Chiang, Spal, Denenstein, Heeger, Miro and
		MacDiarmid}]{chiang1977}
	\bibinfo{author}{Chiang\xfnm[ C.]}, \bibinfo{author}{Spal\xfnm[ R.]},
	\bibinfo{author}{Denenstein\xfnm[ A.]}, \bibinfo{author}{Heeger\xfnm[ A.]},
	\bibinfo{author}{Miro\xfnm[ N.]}, \bibinfo{author}{MacDiarmid\xfnm[ A.]}.
	\newblock \bibinfo{title}{Anomalous electrical properties of linear chain
		mercury compounds}.
	\newblock \bibinfo{journal}{Solid State Commun}
	\bibinfo{year}{1977};\bibinfo{volume}{22}(\bibinfo{number}{5}):\bibinfo{pages}{293}.
	%Type = Article
	\bibitem[{Koteles et~al.(1976)Koteles, Datars, Cutforth and
		Gillespie}]{koteles1976}
	\bibinfo{author}{Koteles\xfnm[ E.]}, \bibinfo{author}{Datars\xfnm[ W.]},
	\bibinfo{author}{Cutforth\xfnm[ B.]}, \bibinfo{author}{Gillespie\xfnm[ R.]}.
	\newblock \bibinfo{title}{Anisotropic optical reflectance of
		{Hg$_{2.91}$SbF$_6$}}.
	\newblock \bibinfo{journal}{Solid State Commun}
	\bibinfo{year}{1976};\bibinfo{volume}{20}(\bibinfo{number}{12}):\bibinfo{pages}{1129}.
	%Type = Article
	\bibitem[{Peebles et~al.(1977)Peebles, Chiang, Cohen, Heeger, Miro and
		MacDiarmid}]{peebles1977}
	\bibinfo{author}{Peebles\xfnm[ D.L.]}, \bibinfo{author}{Chiang\xfnm[ C.]},
	\bibinfo{author}{Cohen\xfnm[ M.J.]}, \bibinfo{author}{Heeger\xfnm[ A.]},
	\bibinfo{author}{Miro\xfnm[ N.]}, \bibinfo{author}{MacDiarmid\xfnm[ A.]}.
	\newblock \bibinfo{title}{Optical properties of linear-chain mercury
		compounds}.
	\newblock \bibinfo{journal}{Phys Rev B}
	\bibinfo{year}{1977};\bibinfo{volume}{15}(\bibinfo{number}{10}):\bibinfo{pages}{4607}.
	%Type = Incollection
	\bibitem[{Brown et~al.(1983)Brown, Datars and Gillespie}]{brown1983}
	\bibinfo{author}{Brown\xfnm[ I.]}, \bibinfo{author}{Datars\xfnm[ W.]},
	\bibinfo{author}{Gillespie\xfnm[ R.]}.
	\newblock \bibinfo{title}{The infinite linear chain compounds
		{Hg$_{3-\delta}$AsF$_6$ and Hg$_{3-\delta}$SbF$_6$}}.
	\newblock In: \bibinfo{editor}{Miller\xfnm[ J.S.]}, editor.
	\bibinfo{booktitle}{Extended Linear Chain Compounds};
	vol.~\bibinfo{volume}{3}. \bibinfo{year}{1983}, p.~\bibinfo{pages}{1}.
	%Type = Article
	\bibitem[{Muts et~al.(2011)Muts, Matar, Rodewald, Vasyli and
		Poettgen}]{muts2011}
	\bibinfo{author}{Muts\xfnm[ I.]}, \bibinfo{author}{Matar\xfnm[ S.F.]},
	\bibinfo{author}{Rodewald\xfnm[ U.C.]}, \bibinfo{author}{Vasyli\xfnm[ Z.]},
	\bibinfo{author}{Poettgen\xfnm[ R.]}.
	\newblock \bibinfo{title}{{SrAu4. 76In1. 24} with {YbMo2Al4}-type structure}.
	\newblock \bibinfo{journal}{Z Naturforsch, B: Chem Sci}
	\bibinfo{year}{2011};\bibinfo{volume}{66}(\bibinfo{number}{10}):\bibinfo{pages}{993}.
	%Type = Article
	\bibitem[{Matar and P{\"o}ttgen(2013)}]{matar2013}
	\bibinfo{author}{Matar\xfnm[ S.F.]}, \bibinfo{author}{P{\"o}ttgen\xfnm[ R.]}.
	\newblock \bibinfo{title}{Electronic structure and bonding in {YTi2Ga4} gallide
		with linear titanium chains and four-bonded gallium atoms}.
	\newblock \bibinfo{journal}{Z Naturforsch, B: Chem Sci}
	\bibinfo{year}{2013};\bibinfo{volume}{68}(\bibinfo{number}{1}):\bibinfo{pages}{23}.
	%Type = Article
	\bibitem[{Gerke et~al.(2013)Gerke, Niehaus, Hoffmann and
		P{\"o}tttgen}]{gerke2013}
	\bibinfo{author}{Gerke\xfnm[ B.]}, \bibinfo{author}{Niehaus\xfnm[ O.]},
	\bibinfo{author}{Hoffmann\xfnm[ R.D.]}, \bibinfo{author}{P{\"o}tttgen\xfnm[
		R.]}.
	\newblock \bibinfo{title}{Infinite linear zinc chains in {AAu4Zn2 (A= Ca, Ce,
			Pr, Nd)} with {YbAl4Mo2} type structure}.
	\newblock \bibinfo{journal}{Z Anorg Allg Chem}
	\bibinfo{year}{2013};\bibinfo{volume}{639}(\bibinfo{number}{14}):\bibinfo{pages}{2575}.
	%Type = Article
	\bibitem[{Tappe et~al.(2013)Tappe, Matar, Schwickert, Winter, Gerke and
		P{\"o}tttgen}]{tappe2013}
	\bibinfo{author}{Tappe\xfnm[ F.]}, \bibinfo{author}{Matar\xfnm[ S.F.]},
	\bibinfo{author}{Schwickert\xfnm[ C.]}, \bibinfo{author}{Winter\xfnm[ F.]},
	\bibinfo{author}{Gerke\xfnm[ B.]}, \bibinfo{author}{P{\"o}tttgen\xfnm[ R.]}.
	\newblock \bibinfo{title}{Linear infinite cadmium chains in {CaAu$_4$Cd$_2$}
		and other intermetallics with {YbMo$_2$Al$_4$}-type structure}.
	\newblock \bibinfo{journal}{Monatsh Chem}
	\bibinfo{year}{2013};\bibinfo{volume}{144}(\bibinfo{number}{6}):\bibinfo{pages}{751}.
	%Type = Article
	\bibitem[{Pokluda et~al.(2015)Pokluda, {\v C}ern{\'y}, {\v S}ob and
		Umeno}]{Pokluda2015}
	\bibinfo{author}{Pokluda\xfnm[ J.]}, \bibinfo{author}{{\v C}ern{\'y}\xfnm[
		M.]}, \bibinfo{author}{{\v S}ob\xfnm[ M.]}, \bibinfo{author}{Umeno\xfnm[
		Y.]}.
	\newblock \bibinfo{title}{Ab initio calculations of mechanical properties:
		Methods and applications}.
	\newblock \bibinfo{journal}{Prog Mater Sci}
	\bibinfo{year}{2015};\bibinfo{volume}{73}:\bibinfo{pages}{127}.
	%Type = Article
	\bibitem[{Feng et~al.(2011{\natexlab{a}})Feng, Xiao, Wan, Qu, Huang, Chen
		et~al.}]{xiao2011}
	\bibinfo{author}{Feng\xfnm[ J.]}, \bibinfo{author}{Xiao\xfnm[ B.]},
	\bibinfo{author}{Wan\xfnm[ C.]}, \bibinfo{author}{Qu\xfnm[ Z.]},
	\bibinfo{author}{Huang\xfnm[ Z.]}, \bibinfo{author}{Chen\xfnm[ J.]}, et~al.
	\newblock \bibinfo{title}{Electronic structure, mechanical properties and
		thermal conductivity of {Ln2Zr2O7 (Ln= La, Pr, Nd, Sm, Eu and Gd)}
		pyrochlore}.
	\newblock \bibinfo{journal}{Acta Mater}
	\bibinfo{year}{2011}{\natexlab{a}};\bibinfo{volume}{59}(\bibinfo{number}{4}):\bibinfo{pages}{1742}.
	%Type = Article
	\bibitem[{Feng et~al.(2011{\natexlab{b}})Feng, Xiao, Chen, Du, Yu and
		Zhou}]{feng2011}
	\bibinfo{author}{Feng\xfnm[ J.]}, \bibinfo{author}{Xiao\xfnm[ B.]},
	\bibinfo{author}{Chen\xfnm[ J.]}, \bibinfo{author}{Du\xfnm[ Y.]},
	\bibinfo{author}{Yu\xfnm[ J.]}, \bibinfo{author}{Zhou\xfnm[ R.]}.
	\newblock \bibinfo{title}{Stability, thermal and mechanical properties of
		{PtxAly} compounds}.
	\newblock \bibinfo{journal}{Mater Des}
	\bibinfo{year}{2011}{\natexlab{b}};\bibinfo{volume}{32}(\bibinfo{number}{6}):\bibinfo{pages}{3231}.
	%Type = Article
	\bibitem[{Feng et~al.(2012)Feng, Xiao, Zhou, Pan and Clarke}]{feng2012}
	\bibinfo{author}{Feng\xfnm[ J.]}, \bibinfo{author}{Xiao\xfnm[ B.]},
	\bibinfo{author}{Zhou\xfnm[ R.]}, \bibinfo{author}{Pan\xfnm[ W.]},
	\bibinfo{author}{Clarke\xfnm[ D.R.]}.
	\newblock \bibinfo{title}{Anisotropic elastic and thermal properties of the
		double perovskite slab--rock salt layer {Ln$_2$SrAl$_2$O$_7$ (Ln = La, Nd,
			Sm, Eu, Gd or Dy)} natural superlattice structure}.
	\newblock \bibinfo{journal}{Acta Mater}
	\bibinfo{year}{2012};\bibinfo{volume}{60}(\bibinfo{number}{8}):\bibinfo{pages}{3380}.
	%Type = Article
	\bibitem[{Feng et~al.(2013)Feng, Xiao, Zhou and Pan}]{feng2013}
	\bibinfo{author}{Feng\xfnm[ J.]}, \bibinfo{author}{Xiao\xfnm[ B.]},
	\bibinfo{author}{Zhou\xfnm[ R.]}, \bibinfo{author}{Pan\xfnm[ W.]}.
	\newblock \bibinfo{title}{Anisotropy in elasticity and thermal conductivity of
		monazite-type {REPO4 (RE= La, Ce, Nd, Sm, Eu and Gd)} from first-principles
		calculations}.
	\newblock \bibinfo{journal}{Acta Mater}
	\bibinfo{year}{2013};\bibinfo{volume}{61}(\bibinfo{number}{19}):\bibinfo{pages}{7364}.
	%Type = Article
	\bibitem[{Sun et~al.(2013)Sun, Gao, Xiao, Li and Wang}]{sun2013}
	\bibinfo{author}{Sun\xfnm[ L.]}, \bibinfo{author}{Gao\xfnm[ Y.]},
	\bibinfo{author}{Xiao\xfnm[ B.]}, \bibinfo{author}{Li\xfnm[ Y.]},
	\bibinfo{author}{Wang\xfnm[ G.]}.
	\newblock \bibinfo{title}{Anisotropic elastic and thermal properties of
		titanium borides by first-principles calculations}.
	\newblock \bibinfo{journal}{J Alloys Compd}
	\bibinfo{year}{2013};\bibinfo{volume}{579}:\bibinfo{pages}{457}.
	%Type = Article
	\bibitem[{Gao et~al.(2014)Gao, Jiang, Zhou and Feng}]{gao2014}
	\bibinfo{author}{Gao\xfnm[ X.]}, \bibinfo{author}{Jiang\xfnm[ Y.]},
	\bibinfo{author}{Zhou\xfnm[ R.]}, \bibinfo{author}{Feng\xfnm[ J.]}.
	\newblock \bibinfo{title}{Stability and elastic properties of {Y-C} binary
		compounds investigated by first principles calculations}.
	\newblock \bibinfo{journal}{J Alloys Compd}
	\bibinfo{year}{2014};\bibinfo{volume}{587}:\bibinfo{pages}{819}.
	%Type = Article
	\bibitem[{Yang et~al.(2016)Yang, Shahid, Zhao, Feng, Wan and Pan}]{yang2016}
	\bibinfo{author}{Yang\xfnm[ J.]}, \bibinfo{author}{Shahid\xfnm[ M.]},
	\bibinfo{author}{Zhao\xfnm[ M.]}, \bibinfo{author}{Feng\xfnm[ J.]},
	\bibinfo{author}{Wan\xfnm[ C.]}, \bibinfo{author}{Pan\xfnm[ W.]}.
	\newblock \bibinfo{title}{Physical properties of {La$_2$B$_2$O$_7$ (B = Zr, Sn,
			Hf and Ge)} pyrochlore: First-principles calculations}.
	\newblock \bibinfo{journal}{J Alloys Compd}
	\bibinfo{year}{2016};\bibinfo{volume}{663}:\bibinfo{pages}{834}.
	%Type = Article
	\bibitem[{Giannozzi et~al.(2009)Giannozzi, Baroni, Bonini, Calandra, Car,
		Cavazzoni et~al.}]{Giannozzi2009}
	\bibinfo{author}{Giannozzi\xfnm[ P.]}, \bibinfo{author}{Baroni\xfnm[ S.]},
	\bibinfo{author}{Bonini\xfnm[ N.]}, \bibinfo{author}{Calandra\xfnm[ M.]},
	\bibinfo{author}{Car\xfnm[ R.]}, \bibinfo{author}{Cavazzoni\xfnm[ C.]},
	et~al.
	\newblock \bibinfo{title}{{QUANTUM ESPRESSO}: a modular and open-source
		software project for quantum simulations of materials}.
	\newblock \bibinfo{journal}{J Phys Condens Matter}
	\bibinfo{year}{2009};\bibinfo{volume}{21}:\bibinfo{pages}{395502}.
	%Type = Article
	\bibitem[{Kohn and Sham(1965)}]{kohn1965}
	\bibinfo{author}{Kohn\xfnm[ W.]}, \bibinfo{author}{Sham\xfnm[ L.J.]}.
	\newblock \bibinfo{title}{Self-consistent equations including exchange and
		correlation effects}.
	\newblock \bibinfo{journal}{Phys Rev}
	\bibinfo{year}{1965};\bibinfo{volume}{140}:\bibinfo{pages}{A1133}.
	%Type = Article
	\bibitem[{Schlipf and Gygi(2015)}]{Schlipf2015}
	\bibinfo{author}{Schlipf\xfnm[ M.]}, \bibinfo{author}{Gygi\xfnm[ F.]}.
	\newblock \bibinfo{title}{Optimization algorithm for the generation of oncv
		pseudopotentials}.
	\newblock \bibinfo{journal}{Comput Phys Commun}
	\bibinfo{year}{2015};\bibinfo{volume}{196}:\bibinfo{pages}{36--44}.
	%Type = Book
	\bibitem[{Gross and Dreizler(2013)}]{gross2013}
	\bibinfo{author}{Gross\xfnm[ E.K.]}, \bibinfo{author}{Dreizler\xfnm[ R.M.]}.
	\newblock \bibinfo{title}{Density functional theory}; vol.
	\bibinfo{volume}{337}.
	\newblock \bibinfo{publisher}{Springer Science \& Business Media};
	\bibinfo{year}{2013}.
	%Type = Article
	\bibitem[{Perdew et~al.(1996)Perdew, Burke and Ernzerhof}]{perdew1996}
	\bibinfo{author}{Perdew\xfnm[ J.P.]}, \bibinfo{author}{Burke\xfnm[ K.]},
	\bibinfo{author}{Ernzerhof\xfnm[ M.]}.
	\newblock \bibinfo{title}{Generalized gradient approximation made simple}.
	\newblock \bibinfo{journal}{Phys Rev Lett}
	\bibinfo{year}{1996};\bibinfo{volume}{77}:\bibinfo{pages}{3865}.
	%Type = Article
	\bibitem[{Monkhorst and Pack(1976)}]{monkhorst1976}
	\bibinfo{author}{Monkhorst\xfnm[ H.J.]}, \bibinfo{author}{Pack\xfnm[ J.D.]}.
	\newblock \bibinfo{title}{Special points for brillouin-zone integrations}.
	\newblock \bibinfo{journal}{Phys Rev B}
	\bibinfo{year}{1976};\bibinfo{volume}{13}:\bibinfo{pages}{5188}.
	%Type = Article
	\bibitem[{Marzari et~al.(1999)Marzari, Vanderbilt, {De Vita} and
		Payne}]{marzari1999}
	\bibinfo{author}{Marzari\xfnm[ N.]}, \bibinfo{author}{Vanderbilt\xfnm[ D.]},
	\bibinfo{author}{{De Vita}\xfnm[ A.]}, \bibinfo{author}{Payne\xfnm[ M.C.]}.
	\newblock \bibinfo{title}{Thermal contraction and disordering of the {Al(110)}
		surface}.
	\newblock \bibinfo{journal}{Phys Rev Lett}
	\bibinfo{year}{1999};\bibinfo{volume}{82}:\bibinfo{pages}{3296}.
	%Type = Article
	\bibitem[{Arias et~al.(1992)Arias, Payne and Joannopoulos}]{arias1992}
	\bibinfo{author}{Arias\xfnm[ T.]}, \bibinfo{author}{Payne\xfnm[ M.]},
	\bibinfo{author}{Joannopoulos\xfnm[ J.]}.
	\newblock \bibinfo{title}{Ab initio molecular-dynamics techniques extended to
		large-length-scale systems}.
	\newblock \bibinfo{journal}{Phys Rev B}
	\bibinfo{year}{1992};\bibinfo{volume}{45}(\bibinfo{number}{4}):\bibinfo{pages}{1538}.
	%Type = Article
	\bibitem[{Golesorkhtabar et~al.(2013)Golesorkhtabar, Pavone, Spitaler, Puschnig
		and Draxl}]{elastic}
	\bibinfo{author}{Golesorkhtabar\xfnm[ R.]}, \bibinfo{author}{Pavone\xfnm[ P.]},
	\bibinfo{author}{Spitaler\xfnm[ J.]}, \bibinfo{author}{Puschnig\xfnm[ P.]},
	\bibinfo{author}{Draxl\xfnm[ C.]}.
	\newblock \bibinfo{title}{Elastic: A tool for calculating second-order elastic
		constants from first principles}.
	\newblock \bibinfo{journal}{Comp Phys Commun}
	\bibinfo{year}{2013};\bibinfo{volume}{184}:\bibinfo{pages}{1861}.
	%Type = Book
	\bibitem[{Kantorovich(2004)}]{kantorovich2004}
	\bibinfo{author}{Kantorovich\xfnm[ L.]}.
	\newblock \bibinfo{title}{Quantum theory of the solid state: an introduction};
	vol. \bibinfo{volume}{136}.
	\newblock \bibinfo{publisher}{Springer Science \& Business Media};
	\bibinfo{year}{2004}.
	%Type = Article
	\bibitem[{Fornasini and Palenzona(1976)}]{fornasini1976}
	\bibinfo{author}{Fornasini\xfnm[ M.]}, \bibinfo{author}{Palenzona\xfnm[ A.]}.
	\newblock \bibinfo{title}{Crystal structure of the ternary {RE Mo2Al4} phases
		{(RE= Gd, Er, Yb)}}.
	\newblock \bibinfo{journal}{J Less-Common Met}
	\bibinfo{year}{1976};\bibinfo{volume}{45}:\bibinfo{pages}{137}.
	%Type = Book
	\bibitem[{Born and Huang(1954)}]{born}
	\bibinfo{author}{Born\xfnm[ M.]}, \bibinfo{author}{Huang\xfnm[ K.]}.
	\newblock \bibinfo{title}{Dynamical Theory of Crystal Lattices}.
	\newblock \bibinfo{address}{London}: \bibinfo{publisher}{Oxford University
		Press}; \bibinfo{year}{1954}.
	%Type = Article
	\bibitem[{Mouhat and Coudert(2014)}]{mouhat2014}
	\bibinfo{author}{Mouhat\xfnm[ F.]}, \bibinfo{author}{Coudert\xfnm[ F.X.]}.
	\newblock \bibinfo{title}{Necessary and sufficient elastic stability conditions
		in various crystal systems}.
	\newblock \bibinfo{journal}{Phys Rev B}
	\bibinfo{year}{2014};\bibinfo{volume}{90}:\bibinfo{pages}{224104}.
	%Type = Article
	\bibitem[{Hill(1952)}]{hill1952}
	\bibinfo{author}{Hill\xfnm[ R.]}.
	\newblock \bibinfo{title}{The elastic behaviour of a crystalline aggregate}.
	\newblock \bibinfo{journal}{Proc Phys Soc London, Sect A}
	\bibinfo{year}{1952};\bibinfo{volume}{65}(\bibinfo{number}{5}):\bibinfo{pages}{349}.
	%Type = Book
	\bibitem[{Timoshenko and Goodier(2008)}]{timoshenko2008}
	\bibinfo{author}{Timoshenko\xfnm[ S.]}, \bibinfo{author}{Goodier\xfnm[ J.]}.
	\newblock \bibinfo{title}{Theory of elasticity}.
	\newblock \bibinfo{edition}{3} ed.; \bibinfo{publisher}{McGraw-HIll};
	\bibinfo{year}{2008}.
	%Type = Article
	\bibitem[{Pugh(1954)}]{pugh1954}
	\bibinfo{author}{Pugh\xfnm[ S.]}.
	\newblock \bibinfo{title}{Relations between the elastic moduli and the plastic
		properties of polycrystalline pure metals}.
	\newblock \bibinfo{journal}{LondEdinbDublPhilMag}
	\bibinfo{year}{1954};\bibinfo{volume}{45}(\bibinfo{number}{367}):\bibinfo{pages}{823}.
	%Type = Article
	\bibitem[{Xiao et~al.(2010)Xiao, Feng, Zhou, Xing, Xie, Cheng
		et~al.}]{xiao2010}
	\bibinfo{author}{Xiao\xfnm[ B.]}, \bibinfo{author}{Feng\xfnm[ J.]},
	\bibinfo{author}{Zhou\xfnm[ C.]}, \bibinfo{author}{Xing\xfnm[ J.]},
	\bibinfo{author}{Xie\xfnm[ X.]}, \bibinfo{author}{Cheng\xfnm[ Y.]}, et~al.
	\newblock \bibinfo{title}{The elasticity, bond hardness and thermodynamic
		properties of {X2B (X= Cr, Mn, Fe, Co, Ni, Mo, W)} investigated by {DFT}
		theory}.
	\newblock \bibinfo{journal}{Physica B}
	\bibinfo{year}{2010};\bibinfo{volume}{405}(\bibinfo{number}{5}):\bibinfo{pages}{1274}.
	%Type = Article
	\bibitem[{Duan et~al.(2014)Duan, Sun, Peng and Zhou}]{duan2014}
	\bibinfo{author}{Duan\xfnm[ Y.]}, \bibinfo{author}{Sun\xfnm[ Y.]},
	\bibinfo{author}{Peng\xfnm[ M.]}, \bibinfo{author}{Zhou\xfnm[ S.]}.
	\newblock \bibinfo{title}{Anisotropic elastic properties of the {Ca-Pb}
		compounds}.
	\newblock \bibinfo{journal}{J Alloys Compd}
	\bibinfo{year}{2014};\bibinfo{volume}{595}:\bibinfo{pages}{14}.
	%Type = Article
	\bibitem[{Chen et~al.(2011)Chen, Niu, Li and Li}]{chen2011}
	\bibinfo{author}{Chen\xfnm[ X.Q.]}, \bibinfo{author}{Niu\xfnm[ H.]},
	\bibinfo{author}{Li\xfnm[ D.]}, \bibinfo{author}{Li\xfnm[ Y.]}.
	\newblock \bibinfo{title}{Modeling hardness of polycrystalline materials and
		bulk metallic glasses}.
	\newblock \bibinfo{journal}{Intermetallics}
	\bibinfo{year}{2011};\bibinfo{volume}{19}(\bibinfo{number}{9}):\bibinfo{pages}{12755}.
	%Type = Article
	\bibitem[{Cazzani and Rovati(2005)}]{cazzani2005}
	\bibinfo{author}{Cazzani\xfnm[ A.]}, \bibinfo{author}{Rovati\xfnm[ M.]}.
	\newblock \bibinfo{title}{Extrema of {Young's} modulus for elastic solids with
		tetragonal symmetry}.
	\newblock \bibinfo{journal}{Int J Solids Struct}
	\bibinfo{year}{2005};\bibinfo{volume}{42}:\bibinfo{pages}{5057}.
	%Type = Book
	\bibitem[{Zener(1948)}]{zener1948}
	\bibinfo{author}{Zener\xfnm[ C.]}.
	\newblock \bibinfo{title}{Elasticity and anelasticity of metals}.
	\newblock \bibinfo{publisher}{University of Chicago press};
	\bibinfo{year}{1948}.
	%Type = Article
	\bibitem[{Chung and Buessem(1967)}]{chung1967}
	\bibinfo{author}{Chung\xfnm[ D.]}, \bibinfo{author}{Buessem\xfnm[ W.]}.
	\newblock \bibinfo{title}{The elastic anisotropy of crystals}.
	\newblock \bibinfo{journal}{J Appl Phys}
	\bibinfo{year}{1967};\bibinfo{volume}{38}(\bibinfo{number}{5}):\bibinfo{pages}{2010}.
	%Type = Article
	\bibitem[{Ledbetter and Migliori(2006)}]{ledbetter2006}
	\bibinfo{author}{Ledbetter\xfnm[ H.]}, \bibinfo{author}{Migliori\xfnm[ A.]}.
	\newblock \bibinfo{title}{A general elastic-anisotropy measure}.
	\newblock \bibinfo{journal}{J Appl Phys}
	\bibinfo{year}{2006};\bibinfo{volume}{100}(\bibinfo{number}{6}):\bibinfo{pages}{063516}.
	%Type = Article
	\bibitem[{Ranganathan and Ostoja-Starzewski(2008)}]{ranganathan2008}
	\bibinfo{author}{Ranganathan\xfnm[ S.I.]},
	\bibinfo{author}{Ostoja-Starzewski\xfnm[ M.]}.
	\newblock \bibinfo{title}{Universal elastic anisotropy index}.
	\newblock \bibinfo{journal}{Phys Rev Lett}
	\bibinfo{year}{2008};\bibinfo{volume}{101}(\bibinfo{number}{5}):\bibinfo{pages}{055504}.
	%Type = Article
	\bibitem[{Hayes and Shuvalov(1998)}]{Hayes1998}
	\bibinfo{author}{Hayes\xfnm[ M.]}, \bibinfo{author}{Shuvalov\xfnm[ A.]}.
	\newblock \bibinfo{title}{On the extreme values of {Young's} modulus, the shear
		modulus, and {Poisson's} ratio for cubic materials}.
	\newblock \bibinfo{journal}{Trans ASME}
	\bibinfo{year}{1998};\bibinfo{volume}{65}:\bibinfo{pages}{786}.
	%Type = Book
	\bibitem[{Nye(1985)}]{nye1985}
	\bibinfo{author}{Nye\xfnm[ J.F.]}.
	\newblock \bibinfo{title}{Physical properties of crystals: their representation
		by tensors and matrices}.
	\newblock \bibinfo{publisher}{Oxford university press}; \bibinfo{year}{1985}.
	%Type = Article
	\bibitem[{Goldstein et~al.(2015)Goldstein, Gorodtsov and
		Lisovenko}]{Goldstein2015}
	\bibinfo{author}{Goldstein\xfnm[ R.V.]}, \bibinfo{author}{Gorodtsov\xfnm[
		V.A.]}, \bibinfo{author}{Lisovenko\xfnm[ D.S.]}.
	\newblock \bibinfo{title}{Young's modulus and {P}oisson's ratio for
		seven-constant tetragonal crystals and nano/microtubes}.
	\newblock \bibinfo{journal}{Phys Mesomech}
	\bibinfo{year}{2015};\bibinfo{volume}{18}:\bibinfo{pages}{213}.
	%Type = Article
	\bibitem[{Gunton and Saunders(1972)}]{Gunton1972}
	\bibinfo{author}{Gunton\xfnm[ D.J.]}, \bibinfo{author}{Saunders\xfnm[ G.A.]}.
	\newblock \bibinfo{title}{The young's modulus and {Poisson's} ratio of arsenic,
		antimony, and bismuth}.
	\newblock \bibinfo{journal}{J Mater Sci}
	\bibinfo{year}{1972};\bibinfo{volume}{7}:\bibinfo{pages}{1061}.
	%Type = Article
	\bibitem[{Ballato(1996)}]{ballato96}
	\bibinfo{author}{Ballato\xfnm[ A.]}.
	\newblock \bibinfo{title}{Poisson's ratio for tetragonal, hexagonal, and cubic
		crystals}.
	\newblock \bibinfo{journal}{IEEE Trans Ultrason Ferroelectr Freq Control}
	\bibinfo{year}{1996};\bibinfo{volume}{43}:\bibinfo{pages}{56}.
	%Type = Article
	\bibitem[{Brugger(1965)}]{brugger1965}
	\bibinfo{author}{Brugger\xfnm[ K.]}.
	\newblock \bibinfo{title}{Pure modes for elastic waves in crystals}.
	\newblock \bibinfo{journal}{J Appl Phys}
	\bibinfo{year}{1965};\bibinfo{volume}{36}:\bibinfo{pages}{759}.
	%Type = Article
	\bibitem[{Cahill and Pohl(1988)}]{Cahill1988}
	\bibinfo{author}{Cahill\xfnm[ D.G.]}, \bibinfo{author}{Pohl\xfnm[ R.]}.
	\newblock \bibinfo{title}{Lattice vibrations and heat transport in crystals and
		glasses}.
	\newblock \bibinfo{journal}{Annu Rev Phys Chem}
	\bibinfo{year}{1988};\bibinfo{volume}{39}:\bibinfo{pages}{93--121}.
	%Type = Article
	\bibitem[{McMillan(1968)}]{mcmillan1968}
	\bibinfo{author}{McMillan\xfnm[ W.]}.
	\newblock \bibinfo{title}{Transition temperature of strong-coupled
		superconductors}.
	\newblock \bibinfo{journal}{Phys Rev}
	\bibinfo{year}{1968};\bibinfo{volume}{167}:\bibinfo{pages}{331}.
	%Type = Article
	\bibitem[{Moruzzi et~al.(1988)Moruzzi, Janak and Schwarz}]{Moruzzi1988}
	\bibinfo{author}{Moruzzi\xfnm[ V.L.]}, \bibinfo{author}{Janak\xfnm[ J.F.]},
	\bibinfo{author}{Schwarz\xfnm[ K.]}.
	\newblock \bibinfo{title}{Calculated thermal properties of metals}.
	\newblock \bibinfo{journal}{Phys Rev B}
	\bibinfo{year}{1988};\bibinfo{volume}{37}:\bibinfo{pages}{790--799}.
	%Type = Article
	\bibitem[{Ma et~al.(2015)Ma, Grabowski, K{\"o}rmann, Neugebauer and
		Raabe}]{Duancheng2015}
	\bibinfo{author}{Ma\xfnm[ D.]}, \bibinfo{author}{Grabowski\xfnm[ B.]},
	\bibinfo{author}{K{\"o}rmann\xfnm[ F.]}, \bibinfo{author}{Neugebauer\xfnm[
		J.]}, \bibinfo{author}{Raabe\xfnm[ D.]}.
	\newblock \bibinfo{title}{Ab initio thermodynamics of the {CoCrFeMnNi} high
		entropy alloy: Importance of entropy contributions beyond the configurational
		one}.
	\newblock \bibinfo{journal}{Acta Mater}
	\bibinfo{year}{2015};\bibinfo{volume}{100}:\bibinfo{pages}{90}.
	%Type = Article
	\bibitem[{Liu et~al.(2015)Liu, VanLeeuwen, Shang, Du and Liu}]{Liu2015}
	\bibinfo{author}{Liu\xfnm[ X.L.]}, \bibinfo{author}{VanLeeuwen\xfnm[ B.K.]},
	\bibinfo{author}{Shang\xfnm[ S.L.]}, \bibinfo{author}{Du\xfnm[ Y.]},
	\bibinfo{author}{Liu\xfnm[ Z.K.]}.
	\newblock \bibinfo{title}{On the scaling factor in {D}ebye--{G}r{\"u}neisen
		model: A case study of the {Mg--Zn} binary system}.
	\newblock \bibinfo{journal}{Comput Mater Sci}
	\bibinfo{year}{2015};\bibinfo{volume}{98}:\bibinfo{pages}{34}.
	%Type = Article
	\bibitem[{Zhang et~al.(2018)Zhang, Grabowski, Hickel and
		Neugebauer}]{Zhang2018}
	\bibinfo{author}{Zhang\xfnm[ X.]}, \bibinfo{author}{Grabowski\xfnm[ B.]},
	\bibinfo{author}{Hickel\xfnm[ T.]}, \bibinfo{author}{Neugebauer\xfnm[ J.]}.
	\newblock \bibinfo{title}{Calculating free energies of point defects from ab
		initio}.
	\newblock \bibinfo{journal}{Comput Mater Sci}
	\bibinfo{year}{2018};\bibinfo{volume}{148}:\bibinfo{pages}{249--259}.
	%Type = Article
	\bibitem[{Zhang et~al.(2017)Zhang, Grabowski, K\"ormann, Freysoldt and
		Neugebauer}]{Zhang2017}
	\bibinfo{author}{Zhang\xfnm[ X.]}, \bibinfo{author}{Grabowski\xfnm[ B.]},
	\bibinfo{author}{K\"ormann\xfnm[ F.]}, \bibinfo{author}{Freysoldt\xfnm[ C.]},
	\bibinfo{author}{Neugebauer\xfnm[ J.]}.
	\newblock \bibinfo{title}{Accurate electronic free energies of the 3d, 4d, and
		5d transition metals at high temperatures}.
	\newblock \bibinfo{journal}{Phys Rev B}
	\bibinfo{year}{2017};\bibinfo{volume}{95}:\bibinfo{pages}{165126}.
	%Type = Article
	\bibitem[{Gupta et~al.(2017)Gupta, Kavakbasi, Dutta, Grabowski, Peterlechner,
		Hickel et~al.}]{Gupta2017}
	\bibinfo{author}{Gupta\xfnm[ A.]}, \bibinfo{author}{Kavakbasi\xfnm[ B.T.]},
	\bibinfo{author}{Dutta\xfnm[ B.]}, \bibinfo{author}{Grabowski\xfnm[ B.]},
	\bibinfo{author}{Peterlechner\xfnm[ M.]}, \bibinfo{author}{Hickel\xfnm[ T.]},
	et~al.
	\newblock \bibinfo{title}{Low-temperature features in the heat capacity of
		unary metals and intermetallics for the example of bulk aluminum and
		al$_3$sc}.
	\newblock \bibinfo{journal}{Phys Rev B}
	\bibinfo{year}{2017};\bibinfo{volume}{95}:\bibinfo{pages}{094307}.
	%Type = Article
	\bibitem[{Glensk et~al.(2015)Glensk, Grabowski, Hickel and
		Neugebauer}]{Glensk2015}
	\bibinfo{author}{Glensk\xfnm[ A.]}, \bibinfo{author}{Grabowski\xfnm[ B.]},
	\bibinfo{author}{Hickel\xfnm[ T.]}, \bibinfo{author}{Neugebauer\xfnm[ J.]}.
	\newblock \bibinfo{title}{Understanding anharmonicity in fcc materials: From
		its origin to ab initio strategies beyond the quasiharmonic approximation}.
	\newblock \bibinfo{journal}{Phys Rev Lett}
	\bibinfo{year}{2015};\bibinfo{volume}{114}:\bibinfo{pages}{195901}.
	%Type = Article
	\bibitem[{Birch(1947)}]{birch1947}
	\bibinfo{author}{Birch\xfnm[ F.]}.
	\newblock \bibinfo{title}{Finite elastic strain of cubic crystals}.
	\newblock \bibinfo{journal}{Phys Rev}
	\bibinfo{year}{1947};\bibinfo{volume}{71}:\bibinfo{pages}{809}.
	%Type = Book
	\bibitem[{Poirier(2000)}]{poirier2000}
	\bibinfo{author}{Poirier\xfnm[ J.P.]}.
	\newblock \bibinfo{title}{Introduction to the Physics of the Earth's interior}.
	\newblock \bibinfo{edition}{2nd} ed.; \bibinfo{publisher}{Cambridge University
		Press}; \bibinfo{year}{2000}.
	
\end{thebibliography}
\end{document}